\documentstyle[preprint,eqsecnum,oldlfont,aps]{revtex}
\begin{document}
\draft
\title{Dynamics of heteropolymers in dilute solution: effective
       equation of motion and relaxation spectrum}
\author{Jiunn-Ren Roan and E. I. Shakhnovich}
\address{Chemistry Department, 12 Oxford Street, Cambridge,
    MA 02138, USA}
\date{\today}
\maketitle
\begin{abstract}
    The dynamics of a heteropolymer chain in solution
    is studied in the limit of long chain length.
    Using functional integral representation we derive
    an effective equation of motion, in which the heterogeneity of
    the chain manifests itself as a time-dependent excluded volume
    effect. At the mean field level, the heteropolymer chain is therefore
    dynamically equivalent to a homopolymer chain with both
    time-independent and time-dependent excluded volume effects.
    The perturbed relaxation spectrum is also calculated. We find
    that heterogeneity also renormalizes the relaxation spectrum.
    However, we find,  to the lowest order in heterogeneity, that
    the relaxation spectrum does not exhibit any dynamic freezing, at
    the point when static (equilibrium) ``freezing'' transition
    occurs in heteropolymer. Namely, the breaking of fluctuation-dissipation
    theorem (FDT) proposed for spin glass dynamics does not have dynamic
    effect in heteropolymer, as far as relaxation spectrum is
    concerned.  The implication of
    this result is discussed.
\end{abstract}

\narrowtext

\section{Introduction}

   Although systems
   without quenched randomness usually provide useful information
   and serve as the first step of our understanding of condensed phase,
   there are situations in which quenched randomness brings entirely
   new phenomena. Flux line pinning in type II
   superconductors, critical to the practical application of
   superconductors in high field magnets, is a manifestation
   of the embedded quenched impurities \cite{super}. In the mixture of
   $^3$He and $^4$He, quenched randomness introduced by porous
   media (e.g. aerogels) shifts the tricritical point and extends
   the range of superfluid phase \cite{helium}. The interesting
   properties (e.g. the onset of remanence effects below a freezing
   temperature) of dilute magnets, modeled by ``spin glass'' \cite{spin},
   are consequences of the magnetic impurities. These are just a
   few examples of the new phenomena brought by quenched randomness.
   Study of these phenomena constitutes an important part of modern
   condensed phase physics.

   Theoretical progress in the study of these systems, especially spin
   glass systems, also has brought impetus to other, somewhat distant, fields.
   For example, it is shown that there is an interesting analogy
   between some models developed for spin glasses and the statistical
   properties of polymers with quenched randomness
   (realized as heterogeneity of the polymer) \cite{bio}.
   This analogy has attracted much attention in recent
   years due to its possible biophysical applications.
   As much work has been focused on the static (thermodynamic)
   consequences of this analogy, we shall discuss in this paper
   its dynamic consequence.

   One of the general methods used to study systems with quenched
   randomness is the so-called replica method \cite{spin,replica}.
   Because the randomness is quenched, thermal equilibrium is
   established without thermally equilibrating with the randomness.
   Therefore, to calculate the thermodynamic quantities for a
   system with quenched randomness one has to calculate the
   average of the free energy over the distribution of
   the randomness,
\begin{equation}
   \langle -k_BT\ln Z\rangle.
\end{equation}
   The mathematical obstacle lies in the difficulty in calculating
   the average of a logarithmic function depending on the randomness.
   Replica method circumvents this obstacle by using the identity
\begin{equation}
   \ln Z=\lim_{n\to 0} {Z^n-1\over n}
\end{equation}

   Replica method has been very useful and successful, though not
   free from controversy. With its success and popularity in spin
   glasses, it is not surprising that the study of heteropolymers
   along the line of the above-mentioned analogy often, if not
   exclusively, uses this method to draw thermodynamic conclusions.

   Besides the replica method, there exist other useful theoretical
   methods to study systems with quenched disorder \cite{spin}.
    Since we are interested in the dynamics of
   heteropolymers,
   among these methods the Langevin dynamic method \cite{spin,dyna}
   is the most relevant one. This method does not use
   the unphysical replica limit $n\to 0$ to calculate thermodynamic
   quantities. In Ref.\ \cite{dyna} it is shown that the functional
   integral formalism developed by Martin, Siggia, and Rose for
   classical statistical dynamics \cite{MSR} allows
   average over the quenched randomness without using the
   replica method. It is also shown that this formalism is capable
   of giving dynamic information as well as statics. Therefore,
   although it generally involves more complicated techniques,
   it is worthwhile to study systems with quenched randomness using
   this method. In spin glasses there have been some studies using
   this approach \cite{spin,Sompolinsky,Kirk:p-spin}. On the side
   of heteropolymers, however, to the best of our knowledge, a
   dynamic study similar to these studies is still lacking.
   While the static analogy between spin glasses and heteropolymers
   has been very useful, a similar dynamic study is necessary if
   one wants to use this analogy to address dynamic issues.

   The purpose of this paper is to study the dynamics of
   single heteropolymer chain, i.e. dilute
   heteropolymer solution. It is well known that the dynamic study
   of polymer solutions, especially dilute solutions, must take into
   account the hydrodynamic effect in order to make the prediction
   experimentally relevant \cite{Doi,Bird,Oono:review}.
   Besides hydrodynamic effect, the heterogeneity of the polymer
   chain presumably introduces further effect. One
   expects that the dynamics of the heteropolymer chain will
   depend on these two effects as well as the usual excluded
   volume effect. Here we investigate the importance of the
   heterogeneity to the chain dynamics.

   The dynamics of dilute polymer solution has been extensively
   studied \cite{Doi,Oono:review}. Dynamic quantities such as
   diffusion constant \cite{Jaga}, time-correlation functions \cite{Schaub},
   intrinsic viscosity \cite{vis}, initial decay rate of the dynamical
   scattering \cite{Lee-Baldwin-Oono}, and relaxation spectrum \cite{spec}
   have been calculated. Among these dynamic quantities we are
   most interested in the relaxation spectrum because, in the
   context of Ref.\ \cite{bio}, i.e. protein dynamics, the relaxation
   spectrum is the most relevant quantity with potential application to
   the protein folding problem. In spite of the close analogy between
   models of spin glasses and heteropolymers at static level \cite{bio},
   we find that this analogy is not as complete at dynamic level.
   While there have been attempts to study heteropolymer dynamics at
   the phenomenological level of the hypothetical dynamics of the
   Random Energy Model (REM) \cite{bwkin} , the present microscopic analysis
   suggests that such analogy may not be justified when dynamic issues
   are tackled.

   The structure of this paper is as follows.
   In Sec.\ \ref{model} we define the model Hamiltonian and the
   dynamic equation of the system. We discuss the physics of the
   model Hamiltonian and the dynamic equation. Assumptions made
   are also addressed. In Sec.\ \ref{mf} we derive
   the mean field equation of motion for the chain in the limit of
   long chain length. This effective equation
   of motion can be further simplified when the characteristic time
   scale of the solvent is much shorter than the characteristic time
   scale of the chain. An effective Hamiltonian results from this
   effective equation. It is shown that the effective Hamiltonian
   contains a memory term, and, hence, a time-dependent excluded volume
   interaction. In Sec.\ \ref{mode} we calculate the perturbed
   relaxation spectrum for the chain, to the lowest order in
   excluded volume interaction, hydrodynamic interaction, and heterogeneity.
   We find that, as long as the model Hamiltonian and dynamic equation
   are valid, the relaxation spectrum does not depend on the final
   equilibrium state of the chain. In other words, in spite of the
   static analogy between spin glass and heteropolymer mentioned above,
   the possible breaking of FDT, due to the degeneracy of the
   ground states, found in the mean field dynamics of spin glass
   \cite{spin,Sompolinsky} does not have corresponding dynamical effect,
   at least to order of our calculation, in the mean field dynamics
   of heteropolymer.
   In Sec.\ \ref{analogy} we discuss the analogy of spin glass and
   heteropolymer, using the well-known correspondence between magnetic system
   and self-avoiding random walk. We will see that,
   when the kinetic equations are compared, it is clear that
   the analogy is not very helpful, because the dynamic
   variables used are completely different and may not be
   relevant to each other.
   In Sec.\ \ref{disc} the physical situations to which our model
   may be applicable are discussed. We also give a brief discussion
   of approximations involved and the limitation of the results.
   In App.\ \ref{Kirkwood} we provide an alternative derivation of
   the conventional (Kirkwood) approach used in the dynamic study of dilute
   polymer solution. App.\ \ref{app:mf} gives the detailed derivation
   of the mean field approximation sketched in Sec.\ \ref{mf}.
   In App.\ \ref{1st} we detail the calculation leading to the result
   in Sec.\ \ref{mode}.

\section{The model}
\label{model}
\subsection{Hamiltonian}

   Consider a heteropolymer chain with conformation specified by
   the coordinates of $N$ monomers, ${{\bf r}}_0,\cdots,{{\bf r}}_{N-1}$.
   The chain can be described by the following Hamiltonian,
\begin{mathletters}
\label{Hamiltonian}
\begin{equation}
  H=H_0+H_R,\label{H}
\end{equation}
\begin{equation}
  H_0\equiv\sum_{i=0}^{N-1}{1\over 2}
  ({{\bf r}}_{i+1}-{{\bf r}}_i)^2,\label{H0}
\end{equation}
\begin{equation}
  H_R\equiv {1\over 2}\sum_{i\neq j}B_{ij}
  U({{\bf r}}_i-{{\bf r}}_j),\label{HR}
\end{equation}
\end{mathletters}
  where $B_{ij}$ is a Gaussian random number centered at $B_0$
  with variance $B^2$ ($B > 0$),
\begin{equation}
  P(B_{ij})={1\over \sqrt{2\pi B^2}}\exp \Bigl[-{(B_{ij}-B_0)^2\over 2B^2}
					 \Bigr],
     \label{PB}
\end{equation}

  In Eqs.\ (\ref{Hamiltonian}) $H_0$ is the Hamiltonian used in the
  bead-spring chain model (Rouse model) \cite{Bird}, where, for
  convenience, all the spring constants have been taken to be unity
  (see below). The chain connectivity of the polymer is described by
  this term. $H_R$ simulates the excluded volume effect. The
  excluded volume interaction $U({{\bf r}}_i-{{\bf r}}_j)$ between
  monomer $i$ and monomer $j$ is Gaussian-modulated by $B_{ij}$. This
  modulation confers the chain heterogeneity. When $B\to 0$ the
  excluded volume interactions are ''monochromatically'' modulated
  by $B_0$ and the chain becomes homogeneous. At this extreme, the
  Hamiltonian, Eq. (\ref{Hamiltonian}), reduces to the usual Edwards
  Hamiltonian \cite{Oono:review},

\begin{equation}
  H_E\equiv \sum_{i=0}^{N-1}{1\over 2}({{\bf r}}_{i+1}-{{\bf r}}_i)^2
	      + {B_0\over 2}\sum_{i\neq j}U({{\bf r}}_i-{{\bf r}}_j)
\label{HE}
\end{equation}
  On the other hand, a nonvanishing $B$
  means that there exist various monomer-monomer interactions,
  and the chain is essentially heterogeneous. Thus $B$ is a measure of
  chain heterogeneity and will be called {\it heterogeneity parameter}.
  $B_0$ is the usual excluded volume parameter measuring the strength
  of excluded volume effect. Presumably, $B$ depends only on the
  heterogeneity (dispersity) of polymer, while $B_0$ is generally a function of
  temperature and solvent quality.

  The excluded volume parameter $B_0$ will be taken to be positive
  throughout this paper. A negative $B_0$ gives attractive
  monomer-monomer interactions, which requires inclusion of
  repulsive three-body interactions in order to render the
  theory stable \cite{Freed:book,Cloizeaux}. To simplify
  the matter, we will confine ourselves to positive $B_0$,
  i.e. a heteropolymer chain in good solvent.

  The conventional use of the excluded volume interaction
  $U({{\bf r}})$ is the hard-core repulsion
  $\delta({{\bf r}})$ \cite{Oono:review,Freed:book,Cloizeaux}.
  To derive the effective equation of motion at the mean field level
  in a more general form, we will not adopt this convention for
  $U({{\bf r}})$ until Sec.\ \ref{mode}, where the perturbed
  relaxation spectrum is calculated. Before Sec.\ \ref{mode}
  the only assumption on $U({{\bf r}})$ will be translational invariance:
  $U({{\bf r}},{{\bf r}}')=U({{\bf r}}-{{\bf r}}')=U({{\bf r}}'-{{\bf r}})$.

  Obviously, the Hamiltonian $H_0$ ignores the possible variation
  of spring constants. Presumably, this variation exists when the
  chain is heterogeneous. Since our concern here is the effect of random
  excluded volume interactions, we do not consider this variation
  in spring constants. With this simplification, by a properly
  chosen length scale, we can always set all spring constants to be
  unity, as is done in	(\ref{H0}).

  The statics of Hamiltonian (\ref{Hamiltonian}) has been studied in
  Refs.\ \cite{bio}, within the context of biopolymer and protein
  folding. These works share a number of techniques with
  spin-glass models. In the past two decades there has been much
  progress in the theories of spin-glass, in both statics and
  dynamics \cite{spin}. However, there has never been any
  systematic microscopic study on the dynamics of
  Hamiltonian (\ref{Hamiltonian}).

\subsection{Dynamics}

  Starting with the Hamiltonian (\ref{Hamiltonian}), for dynamic study,
  we consider the
  coupled Langevin equations proposed by
  Oono and Freed \cite{Oono-Freed:eq}:
\begin{mathletters}
\label{O-F eq}
\begin{equation}
  {\partial\over\partial t}{{\bf r}}_i(t)=-{1\over \zeta_0}
  {\partial H\over\partial{{\bf r}}_i(t)}+g_0{{\bf u}} ({{\bf r}}_i(t),t)
  +{{\bbox{\theta}}_0}_i(t),	  \label{req}
\end{equation}
\begin{equation}
  \langle{{\bbox{\theta}}_0}_i(t)\rangle=0,\qquad
  \langle{{\bbox{\theta}}_0}_i(t){{\bbox{\theta}}_0}_j(t')
  \rangle={2\over\zeta_0}
  \delta(t-t')\delta_{ij}\bbox{1},   \label{rnoise}
\end{equation}
\begin{equation}
  {\partial\over\partial t}{{\bf u}} ({{\bf r}},t)=\nu_0\Delta{{\bf u}}
  ({{\bf r}},t)-{g_0\over\rho_0}\sum_{i=0}^{N-1}{\partial H\over
  \partial{{\bf r}}_i(t)}\delta({{\bf r}}-{{\bf r}}_i(t))-{1\over\rho_0}
  {\bbox{\nabla}} p+{\bbox{\xi}}_0({{\bf r}},t),    \label{ueq}
\end{equation}
\begin{equation}
  \langle{\bbox{\xi}}_0({{\bf r}},t)\rangle=0,\qquad
  \langle{\bbox{\xi}}_0({{\bf r}},t){\bbox{\xi}}_0({{\bf r}}',t')\rangle=
  -{2\eta_0\over\rho_0^2}\Delta\delta({{\bf r}}-{{\bf r}}')
  \delta(t-t')\bbox{1},     \label{unoise}
\end{equation}
\end{mathletters}

   where $\bbox{1}$ is a unit tensor. In these equations the kinetic
   coefficient $\zeta_0^{-1}$ and kinematic viscosity
   $\nu_0$ set the time scales for polymer and solvent molecules,
   respectively. $g_0$ is a measure of the coupling
   strength between monomers and solvent molecules. (We also
   ignore the possible variation of these couplings due to the
   chain heterogeneity.) Langevin noise ${{\bbox{\theta}}_0}_i(t)$
   gives monomer $i$ a random velocity at time $t$. Noise
   ${\bbox{\xi}}_0({{\bf r}},t)$ is the random acceleration of the solvent
   velocity field at position ${{\bf r}}$ and time $t$.
   Dynamic viscosity $\eta_0$ is, as usual, related to the
   solvent density $\rho_0$ by $\eta_0=\rho_0\nu_0$.
   The hydrostatic pressure $p$ ensures the incompressibility
   of the solvent \cite{Landau}. This condition enables us to
   consider only the transverse component of solvent velocity field.
   Since no confusion will be arisen, hereafter, we will denote the
   transverse component of solvent velocity by the same symbol ${{\bf u}}$.

   Eq.\ (\ref{req}) is a relaxation dynamic equation, with mode-coupling
   term $g_0{{\bf u}}$, for the chain. Eq.\ (\ref{ueq}) is the
   Stokes approximation of the Navier-Stokes equation for solvent
   velocity field ${{\bf u}}$, augmented by the forces exerted by
   monomers \cite{Landau}. We will assume that the spatial extension of the
   system is infinite so that we can ignore the boundary condition.
   The problem related to the non-applicability of the Navier-Stokes
   equation in an unbounded space will be also ignored
   \cite{Oono-Baldwin,Beenakker}. Eqs.\ (\ref{O-F eq}) are valid
   as long as the Reynolds number is small.

   In the conventional theory of Brownian motion the effect of
   surrounding solvent molecules is taken into account by the Langevin
   noises. In polymer dynamics the commonly used approach (Kirkwood's
   approach) is essentially based on this viewpoint
   \cite{Doi,Bird,Oono:review}. This approach treats each monomer
   as a point source of frictional force and lumps up the effect of
   solvent molecules in the Langevin noises.
   However, as argued by Oono \cite{Oono:review,Oono:aip},
   this picture is not self-consistent
   because the monomers are not legitimate Brownian particles,
   even if the polymer as a whole can be treated like a Brownian
   object. In contrast to Kirkwood's approach, the coupled Langevin
   equations Eqs.\ (\ref{O-F eq}) proposed
   by Oono and Freed explicitly introduces solvent velocity field.
   In this approach the Langevin noises ${\bbox{\theta}}_0$ and
   ${\bbox{\xi}}_0$ have to be re-interpreted as coming from the
   coarse-graining procedures used to derive these equations, separately,
   from the more fundamental, microscopic, equations, and the coupling
   between polymer and solvent is introduced only after the
   respective coarse-graining procedures have been performed.
   A proper term for these equations is ''kinetic equation''
   \cite{Kawasaki-Gunton}. Therefore, we will simply call
   Eqs.\ (\ref{O-F eq}) {\it Oono-Freed kinetic equation}.

   In the Oono-Freed kinetic equation the energy scale has been chosen
   as $k_BT=1$. The correlation of Langevin noises chosen here, Eqs.\
   (\ref{rnoise}) and (\ref{unoise}) guarantees that the system will
   eventually approach to the equilibrium state determined by
   the statics of Hamiltonian\ (\ref{Hamiltonian}).

   Note that the second term on the right side of Eq.\ (\ref{ueq}) is
   the frictional forces exerted by the monomers, which are point
   sources of frictional force, same as in Kirkwood's approach.
   In fact, there is a close relationship between these two approaches.
   It has been shown that, to the lowest nontrivial order (i.e. to
   the lowest order in the excluded-volume parameter $B_0$ and
   hydrodynamic coupling parameter $g_0$) and within Markovian
   approximation for solvent velocity field, when the solvent velocity
   field is projected out, the Fokker-Planck equation for the kinetic
   equation Eqs.\ (\ref{O-F eq}) reduces to the conventional
   Kirkwood diffusion equation \cite{Oono:review,Lee-Baldwin-Oono}.
   A simplified proof of this reduction is given in App.\ \ref{Kirkwood}.
   These equations are, in this sense, more fundamental than the
   commonly used Kirkwood diffusion equation, upon which most dynamic
   study of dilute polymer solution are built. Therefore, a full
   dynamic study should start from Eqs.\ (\ref{O-F eq}).

   Using the same spirit of the proof in App.\ \ref{Kirkwood}, we first
   project out the velocity field from the kinetic equation. This is
   done by formally solving Eq.\ (\ref{ueq}),
\begin{mathletters}
\label{u's solution}
\begin{eqnarray}
  {{\bf u}}({{\bf r}},t)\,=&&{{\bf u}}_0({{\bf r}},t)+{{\bf u}}_R({{\bf r}},t)
  \\
  {{\bf u}}_0({{\bf r}},t)\equiv &&
      \int_{{\bf k}} e^{i{{\bf k}}{\bbox{\cdot}}{{\bf r}}}{{\bf u}}_0
	       ({{\bf k}},t)
      \nonumber\\
     =&&\int_{{\bf k}} e^{i{{\bf k}}{\bbox{\cdot}}{{\bf r}}}
	       \int_{-\infty}^t dt'
      e^{-\nu_0k^2(t-t')}{{\bf P}}^T({{\bf {\hat{k}}}})
      {\bbox{\cdot}}\biggl[{\bbox{\xi}}_0({{\bf k}},t')-{g_0\over\rho_0}
      \sum_{i=0}^{N-1}{\partial H_0\over\partial{{\bf r}}_i(t')}
      e^{-i{{\bf k}}{\bbox{\cdot}}{{\bf r}}_i(t')}\biggr],   \label{u0}
  \\
  {{\bf u}}_R({{\bf r}},t)\equiv &&
      \int_{{\bf k}} e^{i{{\bf k}}{\bbox{\cdot}}{{\bf r}}}{{\bf u}}_0
		   ({{\bf k}},t)
      \nonumber\\
     =&&\int_{{\bf k}} e^{i{{\bf k}}{\bbox{\cdot}}{{\bf r}}}
		   \int_{-\infty}^t dt'
      e^{-\nu_0k^2(t-t')}{{\bf P}}^T({{\bf {\hat{k}}}})
      {\bbox{\cdot}}\biggl[-{g_0\over\rho_0}
      \sum_{i=0}^{N-1}{\partial H_R\over\partial{{\bf r}}_i(t')}
      e^{-i{{\bf k}}{\bbox{\cdot}}{{\bf r}}_i(t')}\biggr],    \label{uR}
\end{eqnarray}
\end{mathletters}
   In Eqs.\ (\ref{u0}) and (\ref{uR}) ${{\bf P}}^T({{\bf {\hat{k}}}})$
   is the transverse (along ${{\bf {\hat{k}}}}\equiv {{\bf k}}/k$
   direction) projection operator,
\begin{equation}
  {{\bf P}}^T({{\bf {\hat{k}}}})\equiv\bbox{1}-{{\bf {\hat{k}}}}{{\bf
{\hat{k}}}}
\end{equation}
   and
\begin{equation}
  \int_{{\bf k}}\equiv \int {d^d{{\bf k}}\over (2\pi)^d}
\end{equation}
   ($d$ is the spatial dimensionality).
   The Fourier-transformed random acceleration,
\begin{equation}
  {\bbox{\xi}}({{\bf r}},t)\equiv
      \int_{{\bf k}} e^{i{{\bf k}}{\bbox{\cdot}}{{\bf r}}}
	{\bbox{\xi}}({{\bf k}},t),
\end{equation}
   is Gaussian and is correlated according to
\begin{equation}
  \langle{\bbox{\xi}}^\perp({{\bf k}},t){\bbox{\xi}}^\perp
	      ({{\bf k}}',t')\rangle
    ={{\bf P}}^T({{\bf {\hat{k}}}}){2\eta_0\over\rho_0^2}k^2
	      \delta({{\bf k}}+{{\bf k}}')\delta(t-t')
\end{equation}
   where $\perp$ denotes the transverse component,
   ${\bbox{\xi}}^\perp({{\bf k}},t)\equiv{{\bf P}}^T({{\bf {\hat{k}}}})
   {\bbox{\cdot}}{\bbox{\xi}}({{\bf k}},t)$.

   Substituting solution Eq.\ (\ref{u's solution}) into Eq.\ (\ref{req}),
   the equation of motion for the chain becomes
\begin{equation}
  {\partial\over\partial t}{{\bf r}}_i(t)=-{1\over\zeta_0}{\partial H_0\over
  \partial{{\bf r}}_i(t)}+g_0{{\bf u}}_0({{\bf r}}_i(t),t)+
  {{\bbox{\theta}}_0}_i(t)-{1\over\zeta_0}{\partial H_R\over\partial
  {{\bf r}}_i(t)}+g_0{{\bf u}}_R({{\bf r}}_i(t),t),   \label{eq of motion}
\end{equation}
   Eq.\ (\ref{eq of motion}) is the starting equation
   of our dynamic study. Note that this equation has two
   stochatic terms: $g_0{{\bf u}}_0({{\bf r}}_i(t),t)$ which
   contains ${\bbox{\xi}}_0({{\bf k}},t')$, and ${{\bbox{\theta}}_0}_i(t)$.
   In other words, the new Langevin noise is
\begin{equation}
  {{\bbox{\theta}}_0}_i(t)
   +g_0\int_{{\bf k}} e^{i{{\bf k}}{\bbox{\cdot}}{{\bf r}_i(t)}}
	       \int_{-\infty}^t dt'
      e^{-\nu_0k^2(t-t')}{{\bf P}}^T({{\bf {\hat{k}}}})
      {\bbox{\cdot}}{\bbox{\xi}}_0({{\bf k}},t')
\label{newnoise}
\end{equation}
   Obviously, this new Langevin noise is no longer $\delta$-correlated
   in time.

   Eq.\ (\ref{eq of motion}) is much more complicated than its spin-glass
   counterpart \cite{spin,Sompolinsky,Kirk:p-spin}, because of the presence
   of the excluded volume interactions and the hydrodynamic coupling.
   In polymer dynamics, quenched randomness has been considered
   (for example, the mean field dynamics of directed polymer in
   random meida studied by Vilgis \cite{Vilgis}), but the randomness
   is from external random potential rather than chemical dispersity.
   Also, equations similar to Eq.\ (\ref{eq of motion}) in which
   hydrodynamic effect is included have been studied in the context
   of homopolymers
   \cite{Oono:review,Jaga,Schaub,Oono-Freed:eq,Oono:aip,Schaub:1,Schaub:2}
   and two dimensional membranes \cite{Frey}. These studies, however,
   consider simpler situations in which either quenched randomness or
   excluded volume interaction is absent. Eq.\ (\ref{eq of motion})
   considers all three effects: quenched randomness, hydordynamic
   coupling, and excluded volume effect.

\section{Mean field approximation and the effective equation of motion}
\label{mf}
\subsection{Functional-integral representation}
\label{sec:msr}

  The equation of motion, Eq.\ (\ref{eq of motion}), is essentially
  a (nonlinear) stochastic differential equation with colored noise
  (Eq.\ (\ref{newnoise})) \cite{Hanggi}. A standard and convenient
  method to study stochastic differential equations is to use
  functional integral formalism (Martin-Siggia-Rose (MSR) formalism)
  \cite{MSR,Hanggi,Zinn-Justin,Bausch}. The main idea is to write
  down the probability functional for the Langevin noises and make
  change of variables from these noises to dynamical variables.
  The stochastic differential equation itself is treated as a
  constraint, limiting the evolution of probability path. By
  introducing auxiliary fields conjugated to the dynamical variables,
  this constraint, in the form of a $\delta$ functional, can be
  written as a functional integral of these auxiliary fields.
  This leads to a probability functional in terms of the dynamical
  variables and their conjugated auxiliary fields. The auxiliary-field
  technique used in this formalism is similar to the technique used
  in the supersymmetry formalism of stochastic differential equations,
  in which the auxiliary fields are replaced by fermion fields
  \cite{Zinn-Justin,Kurchan}.

   Since the functional integral formalism for stochastic processes is
   already well documented, we would not go into any detail of it.
   Useful references are Refs.\ \cite{MSR,Zinn-Justin,Bausch}.
   (The concise review in Ref.\ \cite{Das-Mazenko} is also helpful.)
   The MSR formalism has been widely used for dynamic study of condensed
   phase systems, for example, liquid-glass transition \cite{Das-Mazenko},
   turbulence \cite{McComb}, spin glass dynamics \cite{Sompolinsky}
   and so on. Nevertheless, in polymer physics it is not often used.
   This is probably because that, in spite of its elegance, no new
   results have been obtained through using this formalism
   \cite{Oono-email,Fred}. The study in Refs.\
   \cite{Jaga,Schaub,Schaub:1,Schaub:2} makes it clear that,
   for practical calculation, the MSR formalism does not achieve,
   at this point, better (higher order) results than the conventional
   approach \cite{Oono:review,Oono:aip}. The mathematical involvement
   makes higher order calculations insurmountable.
   However, for our purposes the MSR method is very useful as it
   allows to carry out quenched averaging in the most natural form
   (see below).

   Using the MSR formalism, the generating functional for
   Eq.\ (\ref{eq of motion}) is presented as:
\begin{equation}
  Z_{R\theta_0\xi_0}=\int \{{\cal D}{{\bf r}}_i(t)\}
		  \{{\cal D}{{\bf {\hat{r}}}}_i(t)\}
		  J(\{{{\bf r}}_i\})e^{L_0+L_R}, \label{Z:MSR}
\end{equation}
\begin{equation}
  L_0=\int_{-\infty}^\infty dt
      \sum_{i=0}^{N-1} i{{\bf {\hat{r}}}}_i(t){\bbox{\cdot}}
      \biggl[{\partial\over\partial t}{{\bf r}}_i(t)+{1\over\zeta_0}
      {\partial H_0\over\partial{{\bf r}}_i(t)}-g_0{{\bf u}}_0
      ({{\bf r}}_i(t),t)-{{\bbox{\theta}}_0}_i(t)\biggr],  \label{L0}
\end{equation}
\begin{equation}
  L_R=\int_{-\infty}^\infty dt
      \sum_{i=0}^{N-1} i{{\bf {\hat{r}}}}_i(t){\bbox{\cdot}}
      \biggl[{1\over\zeta_0}{\partial H_R\over\partial{{\bf r}}_i(t)}
      -g_0{{\bf u}}_R({{\bf r}}_i(t),t)\biggr],  \label{LR}
\end{equation}
   Note that the Langevin noises ${\bbox{\theta}}_0$ and ${\bbox{\xi}}_0$
   are now in $L_0$, while the quenched heterogeneity, represented
   by $B_{ij}$, is in $L_R$. Since ${\bbox{\theta}}_0$,
   ${\bbox{\xi}}_0$ and $B_{ij}$ are not coupled,
   we can take average of $Z_{R\theta_0\xi_0}$ over these random
   quantities separately.

  The Jacobian $J(\{{{\bf r}}_i\})$ in Eq.\ (\ref{Z:MSR})
  is associated with the change of variables from Langevin noises
  to dynamical variables (mentioned above).
  Because of the following two reasons, we will drop the
  Jacobian from our calculation from now on. Firstly,
  the specific form of this Jacobian depends on the time-discretization
  scheme used when the functional integral representation is
  written down \cite{MSR-Jacobian}.
  (For Stratonovich scheme Refs.\ \cite{Graham,Peliti}
  work out the detailed form.) Although it is not uniquely defined,
  one can get correct result provided one consistently uses
  the same discretization scheme \cite{West}. Secondly, and more
  importantly, the Jacobian depends on the dynamical variables
  $\{{{\bf r}}_i\}$, not on the auxiliary variables
  $\{{{\bf {\hat{r}}}}_i\}$. As we
  can see from Eqs.\ (\ref{L0}) and (\ref{LR}) the equation of
  motion, Eq.\ (\ref{eq of motion}), is always coupled to the
  auxiliary fields $\{{{\bf {\hat{r}}}}_i\}$. This will still be true for
  the effective equation of motion that we seek for. Therefore,
  whatever form the Jacobian has, or, whether it also gets
  averaged when we perform averages over $B_{ij}$, ${\bbox{\theta}}_0$
  and ${\bbox{\xi}}_0$ is not important. We do not need it anyway.

  In (\ref{Z:MSR}) only the auxiliary fields $\{{{\bf {\hat{r}}}}_i\}$ are
  introduced. In principle, for the Oono-Freed kinetic equation
  (\ref{O-F eq}), there should also be auxiliary fields
  $\{{{\bf {\hat{u}}}}\}$, as in Ref.\ \cite{Jaga,Schaub,Schaub:1,Schaub:2}.
  Since we are not interested in the solvent velocity fields, ${{\bf u}}$,
  and have eliminated it from the kinetic equation, there is no need to
  introduce an auxiliary variable ${{\bf {\hat{u}}}}$ for ${{\bf u}}$.

  For computational convenience we define three operators,
\begin{equation}
  {\bbox{\Gamma}}_{ij}(t,t')\equiv {g_0^2\over\rho_0}\zeta_0
     \int_{{\bf k}} e^{-\nu_0k^2(t-t')}e^{i{{\bf k}}{\bbox{\cdot}}
     [{{\bf r}}_i(t)-{{\bf r}}_j(t')]}{{\bf P}}^T({{\bf {\hat{k}}}})
			 \label{Gamma}
\end{equation}
  (a symmetric tensor),
\begin{equation}
  a_i(t)\equiv{1\over 2}\int dt'\sum_{j=0}^{N-1}i{{\bf {\hat{r}}}}_j(t')
    {\bbox{\cdot}}\Bigl[\bbox{1}\delta_{ij}\delta(t-t'-\epsilon)\theta(t-t')
    +{\bbox{\Gamma}}_{ji}(t',t)\theta(t'-t)\Bigr]{\bbox{\cdot}}
    {\bbox{\nabla}}_i(t)
			 \label{a}
\end{equation}
   (a scalar differential operator; $\epsilon$ being an infinitesimal
   positive number), and
\begin{equation}
  {\cal O}_{ij}\equiv\int dt
       \Bigl[a_i(t)U({{\bf r}}_i(t)-{{\bf r}}_j(t))
       +a_j(t)U({{\bf r}}_j(t)-{{\bf r}}_i(t))\Bigr]   \label{O}
\end{equation}
   (symmetric in $i$ and $j$).
   Now we can rewrite $L_R$ in a more compact form
\begin{equation}
  L_R=\sum_{i\neq j}{1\over\zeta_0}B_{ij}{\cal O}_{ij}, \label{LR1}
\end{equation}
   The average over $B_{ij}$ can be easily done and gives
\begin{eqnarray}
  Z_{\theta_0\xi_0}=&&\int \{{\cal D}{{\bf r}}_i(t)\}
		\{{\cal D}{{\bf {\hat{r}}}}_i(t)\} e^{L_0}
	     \langle e^{L_R}\rangle_B  \nonumber\\
	  =&&\int \{{\cal D}{{\bf r}}_i(t)\}\}
		  \{{\cal D}{{\bf {\hat{r}}}}_i(t)\} e^{L_0}
	     \exp \biggl[{B^2\over 2\zeta_0^2}\sum_{i\neq j}{\cal O}_{ij}^2
		  +{B_0\over\zeta_0}\sum_{i\neq j}{\cal O}_{ij}\biggr]
\label{Z after B}
\end{eqnarray}

\subsection{Mean field approximation}

   The generating functional obtained in Sec.\ \ref{sec:msr} can be used to
   derive an effective equation of motion in the limit of long chain length
   $N\to\infty$. The same method is used to derive an effective
   equation of motion for spin glass \cite{Sompolinsky,Kirk:p-spin}
   and for directed polymers \cite{Vilgis}. Although the algebra
   involeve here is more complicated, the resulting effective
   equation of motion share the same characteristics as the effective
   equation of motion obtained in Ref.\ \cite{Sompolinsky,Kirk:p-spin}.

   The detailed calculation in the long chain limit can be found in
   App.\ \ref{app:mf}. The result is an effective Lagrangian:
\begin{eqnarray}
  L_e=L_0&&+{B^2\over\zeta_0^2}\int dt dt'd1 d2 d3 d4 \biggl[
	    \Bigl\langle\bar{A}(t,t',1,3)\Bigr\rangle \bar{B}(t,t',2,4)
	   +\Bigl\langle\bar{B}(t,t',2,4)\Bigr\rangle \bar{A}(t,t',1,3)
	\nonumber\\
	&&\qquad\qquad\qquad
	  +2\Bigl\langle\bar{C}(t,t',1,3)\Bigr\rangle \bar{D}(t,t',2,4)\biggr]
	      U(1-2)U(3-4)   \nonumber\\
	&&+{2B_0\over\zeta_0}\int dt dt'd1 d2 d3 d4 \biggl[
	    \Bigl\langle\bar{C}(t,t',1,3)\Bigr\rangle \bar{B}(t,t',2,4)
	\nonumber\\
	&&\qquad\qquad\qquad
	  +\Bigl\langle\bar{B}(t,t',2,4)\Bigr\rangle \bar{C}(t,t',1,3) \biggr]
	     \delta(t-t')U(1-2),   \label{Le}
\end{eqnarray}
   where, for convenience, ${{\bf R}}_1$, ${{\bf R}}_2$, ${{\bf R}}_3$,
   ${{\bf R}}_4$ are written as $1$, $2$, $3$, $4$ respectively. The
   quantities $\bar{A}$, $\bar{B}$, $\bar{C}$, $\bar{D}$ are (see
   App.\ \ref{app:mf})
\begin{mathletters}
\label{abcd}
\begin{eqnarray}
  &&\bar{A}(t,t',{{\bf R}},{{\bf R}}')\equiv\sum_i a_i(t)a_i(t')
      \delta({{\bf r}}_i(t)-{{\bf R}})\delta({{\bf r}}_i(t')-{{\bf R}}')
    \\
  &&\bar{B}(t,t',{{\bf R}},{{\bf R}}')\equiv\sum_i
      \delta({{\bf r}}_i(t)-{{\bf R}})\delta({{\bf r}}_i(t')-{{\bf R}}')
    \\
  &&\bar{C}(t,t',{{\bf R}},{{\bf R}}')\equiv\sum_i a_i(t)
      \delta({{\bf r}}_i(t)-{{\bf R}})\delta({{\bf r}}_i(t')-{{\bf R}}')
    \\
  &&\bar{D}(t,t',{{\bf R}},{{\bf R}}')\equiv\sum_i a_i(t')
      \delta({{\bf r}}_i(t)-{{\bf R}})\delta({{\bf r}}_i(t')-{{\bf R}}')
\end{eqnarray}
\end{mathletters}

   Note that $\bar{B}$ is the dynamic version of the order parameter used in
   Ref.\ \cite{bio}. In spin glass dynamics it is also found that the
   static Edwards-Anderson order parameter should be generalized to a
   time-dependent order parameter whose long time limit gives the static
   order parameter \cite{Sompolinsky,Sompolinsky:order}. Similarly,
   the static order parameter used in Ref.\ \cite{bio} can be defined
   as the long time (equilibrium) limit of this dynamic order parameter
   $\bar{B}$. The physical meaning of the static order parameter is
   explained in Ref.\ \cite{bio}. We will explain later the physical meaning
   of the dynamical order parameter $\bar{B}$.

   A little reflection tells us that, firstly, the terms $\langle\bar{A}
   \rangle\bar{B}$ and $\langle\bar{C}\rangle\bar{B}$ will not appear in the
   final effective equation of motion because they are related to the
   Jacobian when writing the functional integral representation for
   the effective equation of motion. As explained earlier, the Jacobian
   does not play an important role, as far as the effective equation of
   motion is concerned. Hence, we will also ignore these terms from
   now on. Secondly, the term $\langle\bar{B}\rangle\bar{A}$ is related to
   the stochastic part of the effective equation of motion because it
   contains two auxiliary fields outside of the self-consistent average
   bracket $\langle\rangle$. Finally, because of having only one auxiliary
   field outside of the self-consistent average bracket, the terms
   $\langle\bar{C}\rangle  \bar{D}$ and $\langle\bar{B}\rangle\bar{C}$
   are associated with the deterministic part.

   We then take the average over ${\bbox{\theta}}_0$ and ${\bbox{\xi}}_0$.
   Because of their Gaussian character, these averages are easy to carry
   out. Note that these averages give
\begin{equation}
  \Biggl\langle\exp\int dt\sum_ii{{\bf {\hat{r}}}}_i(t){\bbox{\cdot}}
     (-{{\bbox{\theta}}_0}_i(t)\Biggr\rangle_{\theta_0}
  =\exp\biggl[{1\over 2}\int dt dt'\sum_i\sum_j i {{\bf {\hat{r}}}}_i(t)
	{\bbox{\cdot}}
	\langle{{\bbox{\theta}}_0}_i(t){{\bbox{\theta}}_0}_j(t')
	\rangle{\bbox{\cdot}}
	i{{\bf {\hat{r}}}}_j(t') \biggr]
\end{equation}
   and
\begin{eqnarray}
    \Biggl\langle\exp
  &&\int dt\sum_i i{{\bf {\hat{r}}}}_i(t){\bbox{\cdot}}
    \Bigl[-g_0\int_{{\bf k}}\int_{-\infty}^t dt' e^{-\nu_0k^2(t-t')}
    {{\bf P}}^T({{\bf {\hat{k}}}}){\bbox{\cdot}}{\bbox{\xi}}_0({{\bf k}},t')
    e^{i{{\bf k}}{\bbox{\cdot}}{{\bf r}}_i(t)}\Bigr]
    \Biggr\rangle_{\xi_0}
  \nonumber\\
  &&=\exp \Biggl\{\sum_i\sum_j\int dt dt' i{{\bf {\hat{r}}}}_i(t)
     {\bbox{\cdot}}
     \biggl[{g_0^2\over 2}
     \int d\bar{t}d\bar{t}'\int_{{\bf k}}\int_{{{\bf k}}'}
     \theta(t-\bar{t})\theta(t'-\bar{t}')
     e^{-\nu_0k^2(t-\bar{t})}e^{-\nu_0k'^2(t'-\bar{t}')}
  \nonumber\\
  &&\qquad\times e^{i{{\bf k}}{\bbox{\cdot}}{{\bf r}}_i(t)}
     e^{i{{\bf k}}'{\bbox{\cdot}}{{\bf r}}_j(t')}
     \langle{\bbox{\xi}}_0^\perp({{\bf k}},\bar{t})
     {\bbox{\xi}}_0^\perp({{\bf k}}',\bar{t}')\rangle \biggr]
     {\bbox{\cdot}} i{{\bf {\hat{r}}}}_j(t') \Biggr\}
\end{eqnarray}
   in $\langle\exp L_0\rangle_{\theta_0\xi_0}$ and contribute additional
   terms to the random part of the effective equation of motion, which
   is the term $\langle\bar{B}\rangle\bar{A}$ in Eq.\ (\ref{Le}).
   Bearing this in mind, we can easily write down the effective equation
   of motion.

\subsection{Effective equation of motion and Markov approximation}
\subsubsection{Effective equation of motion}
\label{sec:eff}

   It is a simple matter to read off the effective equation of motion
   after the average over Langevin noises is taken. We can explicitly
   write down the expression for $\langle\bar{B}\rangle\bar{A}$ in Eq.\
   (\ref{Le}) and attribute terms involving tensor ${\bbox{\Gamma}}_{ij}$
   to renormalized noise ${\bbox{\xi}}$ and the rest terms to renormalized
   noise ${\bbox{\theta}}$. (See Eqs.\ (\ref{renorm theta}) and
   (\ref{renorm xi}) below.). Because the algebra is straightforward and
   slightly long, we will not write down the details here. The result is
\begin{equation}
  {\partial\over\partial t}{{\bf r}}_i(t)=-{1\over\zeta_0}
  {\partial H_0\over\partial{{\bf r}}_i(t)}+g_0{\cal J}_0
  -{2B^2\over\zeta_0^2}{\cal J}_1-{2B_0\over
  \zeta_0}{\cal J}_2+{\bbox{\theta}}_i(t), \label{eff0}
\label{bare-eff}
\end{equation}
  where
\begin{equation}
  {\cal J}_0
     =\int_{{\bf k}} e^{i{{\bf k}}{\bbox{\cdot}}{{\bf r}}}\int_{-\infty}^t dt'
      e^{-\nu_0k^2(t-t')}{{\bf P}}^T({{\bf {\hat{k}}}})
      {\bbox{\cdot}}\biggl[{\bbox{\xi}}_0({{\bf k}},t')-{g_0\over\rho_0}
      \sum_{i=0}^{N-1}{\partial H_0\over\partial{{\bf r}}_i(t')}
      e^{-i{{\bf k}}{\bbox{\cdot}}{{\bf r}}_i(t')}\biggr], \label{J0}
\end{equation}
   is a vector field similar to the original solvent velocity field Eq.\
   (\ref{u0}). The renormalized Langevin noises are still Gaussian,
   with variances
\begin{equation}
  \Bigl\langle{\bbox{\theta}}_i(t){\bbox{\theta}}_j(t')\Bigr\rangle=
  \Bigl\langle{{\bbox{\theta}}_0}_i(t){{\bbox{\theta}}_0}_j(t')\Bigr\rangle +
  {B^2\over 2\zeta_0^2}{\cal J}_3, \label{renorm theta}
\end{equation}
  and
\begin{equation}
  \Bigl\langle{\bbox{\xi}}^\perp({{\bf k}},t)
	 {\bbox{\xi}}^\perp({{\bf k}}',t')\Bigr\rangle=
  \Bigl\langle{\bbox{\xi}}_0^\perp({{\bf k}},t)
	 {\bbox{\xi}}_0^\perp({{\bf k}}',t')\Bigr\rangle +
  {B^2g_0^2\over 2\rho_0^2}{\cal J}_4, \label{renorm xi}
\end{equation}
   In these equations, for convenience, we
   have defined vectorial quantities ${\cal J}_1$ and ${\cal J}_2$
   and tensorial quantities ${\cal J}_3$ and ${\cal J}_4$. Their
   expressions are
\begin{eqnarray}
 {\cal J}_1&&={1\over 2}\int dt' d1 d2
    \Bigl\langle\bar{D}(t,t',1,2)\Bigr\rangle
    {\bbox{\nabla}}_i(t)U({{\bf r}}_i(t)-1)U({{\bf r}}_i(t')
    -2)   \nonumber\\
    &&\,+{1\over 2}\sum_i\int dt' d\bar{t}d1 d2
    \Bigl\langle\bar{D}(t',\bar{t},1,2)\Bigr\rangle
    \theta(t-t'){\bbox{\Gamma}}_{ij}(t,t'){\bbox{\cdot}}{\bbox{\nabla}}_j(t')
    U({{\bf r}}_j(t')-1)U({{\bf r}}_j(\bar{t})-2)
\end{eqnarray}
\begin{eqnarray}
  {\cal J}_2=&&{1\over 2}\int d1 d2
    \Bigl\langle\bar{B}(t,t,1,2)\Bigr\rangle
    {\bbox{\nabla}}_i(t)U({{\bf r}}_i(t)-1)
    \nonumber\\
    &&+{1\over 2}\sum_i\int dt' d1 d2
    \Bigl\langle\bar{B}(t',t',1,2)\Bigr\rangle
    \theta(t-t'){\bbox{\Gamma}}_{ij}(t,t'){\bbox{\cdot}}{\bbox{\nabla}}_j(t')
    U({{\bf r}}_j(t')-1)
\end{eqnarray}
\begin{eqnarray}
  {\cal J}_3=&&\delta_{ij}\int d1 d2
    \Bigl\langle\bar{B}(t,t',1,2)\Bigr\rangle
    {\bbox{\nabla}}_i(t){\bbox{\nabla}}_i(t')
    U({{\bf r}}_i(t)-1)U({{\bf r}}_i(t')-2)
    \nonumber\\
    &&+2\int d\bar{t} d1 d2
    \Bigl\langle\bar{B}(\bar{t},t',1,2)\Bigr\rangle
    \theta(t-\bar{t}){\bbox{\Gamma}}_{ij}(t,\bar{t})
    {\bbox{\cdot}}{\bbox{\nabla}}_j(\bar{t}){\bbox{\nabla}}_j(t')
    U({{\bf r}}_j(\bar{t})-1)U({{\bf r}}_j(t')-2)
    \nonumber\\
   =&&\delta_{ij}\int d1 d2
    \Bigl\langle\bar{B}(t,t',1,2)\Bigr\rangle
    {\bbox{\nabla}}_i(t){\bbox{\nabla}}_i(t')
    U({{\bf r}}_i(t)-1)U({{\bf r}}_i(t')-2)
    \nonumber\\
    &&+2\int d\bar{t} d1 d2
    \Bigl\langle\bar{B}(t,\bar{t},1,2)\Bigr\rangle
    \theta(t'-\bar{t}){\bbox{\nabla}}_i(t){\bbox{\Gamma}}_{ji}
    (t',\bar{t}){\bbox{\cdot}}{\bbox{\nabla}}_i(\bar{t})
    U({{\bf r}}_i(\bar{t})-1)U({{\bf r}}_i(\bar{t})-2)
\end{eqnarray}
\begin{eqnarray}
  {\cal J}_4&&=\sum_i\int d1 d2
    \Bigl\langle\bar{B}(t,t',1,2)\Bigr\rangle
       e^{-i{{\bf k}}{\bbox{\cdot}}{{\bf r}}_j(t)}e^{-i{{\bf k}}'
		  {\bbox{\cdot}}{{\bf r}}_j(t')}
       \nonumber\\
     &&\qquad\qquad\times{{\bf P}}^T({{\bf {\hat{k}}}})
       {\bbox{\cdot}}{\bbox{\nabla}}_j(t)
       {{\bf P}}^T({{\bf {\hat{k}}}}'){\bbox{\cdot}}{\bbox{\nabla}}_j(t')
       U({{\bf r}}_j(t)-1)U({{\bf r}}_j(t')-2)
\end{eqnarray}
   (We have suppressed all the possible dependence of these ${\cal J}$'s on
   time, monomer index, and so on.)

   Note that, although Eq.\ (\ref{J0}) is essentially the same as
   Eq.\ (\ref{u0}), it would be incorrect to write a separate
   Navier-Stokes equation, similar to Eq.\ (\ref{ueq}), for
   ${\cal J}_0$, in which the full Hamiltonian is replaced by $H_0$,
   and claim that when the average over the quenched randomness is
   performed the effective equation of motion for solvent velocity
   is not changed. Recall that we have eliminated the solvent velocity
   field before the average is performed. In priciple, one can explicitly
   introduce an auxiliary field for the solvent velocity and perform
   the average over the quenched randomness to find effective equations
   of motion for both polymer conformation and solvent velocity.
   The effective equation of motion for solvent velocity field is
   presumably different from the original Navier-Stokes equation,
   though we expect that we would get the same equation as Eq.\
   (\ref{bare-eff}) once the velocity field is projected out.
   As we have mentioned in Sec.\ \ref{model}, our concern here
   is the equation of motion of the chain. Therefore, we
   will make no attempt on deriving an effective Oono-Freed
   kinetic equation for solvent velocity field.

   Following the conventional interpretation of MSR formalism
   \cite{MSR,Bausch}, time-correlation functions of dynamical
   variables $\{{{\bf r}}_i\}$ and their auxiliary fields
   $\{{{\bf {\hat{r}}}}_i\}$ are response functions of the system
   to the perturbation associated with $\{{{\bf {\hat{r}}}}_i\}$.
   Causality requires that these response functions should be vanishing
   if any of the time variables of the auxiliary fields
   $\{{{\bf {\hat{r}}}}_i\}$ is larger than all the time variables of
   the dynamical variables $\{{{\bf r}}_i\}$. Imposing this causality
   requirement upon Eqs.\ (\ref{abcd}) we can show that causality
   is also respected in the effective equation of motion, Eq.\ (\ref{eff0}).

\subsubsection{Markov approximation}

   The essential ingredients of the proof given in App.\ \ref{Kirkwood}
   are (1) elimination of solvent velocity field, and (2)
   Markov approximation. Physically, (1) means average over the
   solvent velocity and (2) means that the characteristic time
   scale for solvent dynamics is much shorter than the characteristic
   time scale for the motion of the chain.

   We adopt the same spirit of App.\ \ref{Kirkwood}. As the
   velocity field has been eliminated, we now apply the Markov
   approximation to further simplify the effective equation of motion.
   The Markov approximation is
\begin{equation}
  e^{-\nu_0k^2(t-t')} \to {2\over\nu_0 k^2}\delta(t-t')
\end{equation}
   Within this approximation, the Oseen tensor
\begin{equation}
  {{\bf T}}({{\bf r}},{{\bf r}}')\equiv\int_{{{\bf k}}}{1\over\eta_0 k^2}
	   {{\bf P}}^T({{\bf {\hat{k}}}}) e^{i{{\bf k}}{\bbox{\cdot}}
	   ({{\bf r}}-{{\bf r}}')}
\end{equation}
   appears, for example,
\begin{equation}
  {\bbox{\Gamma}}_{ij}(t,t')=2g_0^2\zeta_0\delta(t-t'){{\bf T}}
    ({{\bf r}}_i(t),{{\bf r}}_j(t))
\end{equation}
   Using Markov approximation we can largely simplify the effective
   equation of motion.

   Since we seek the result only to the lowest nontrivial order, that
   is, to the lowest order in $B$, $g_0$, and $B_0$, we can simplfy
   the expressions of ${\cal J}$'s by dropping all the higher order
   terms. For convenience, we define
\begin{equation}
  \bar\rho({{\bf R}},t)\equiv\Bigl\langle\sum_i
     \delta({{\bf r}}_i(t)-{{\bf R}})\Bigr\rangle
\end{equation}
\begin{equation}
  \bbox{G}(t,t',1,2)\equiv\Bigl\langle\sum_i
    \delta({{\bf r}}_i(t)-1)i{{\bf {\hat{r}}}}_i(t')
    \delta({{\bf r}}_i(t')-2)\Bigr\rangle
\end{equation}
\begin{equation}
  C(t,t',1,2)\equiv\Bigl\langle\sum_i
      \delta({{\bf r}}_i(t)-1)
      \delta({{\bf r}}_i(t')-2)\Bigr\rangle
\end{equation}
   where the bracket is the mean-field average defined by
   Eq.\ (\ref{average}) in App.\ \ref{app:mf}.
   The mean-field kinetic equation, Eq.\ (\ref{bare-eff}),
   then reduces to
\begin{equation}
  {\partial\over\partial t}{{\bf r}}_i(t)=-{1\over\zeta_0}
     {\partial H_{eff}\over\partial{{\bf r}}_i(t)}-g_0^2\sum_j
     {{\bf T}}({{\bf r}}_i,{{\bf r}}_j){\bbox{\cdot}}
     {\partial H_{eff}\over\partial{{\bf r}}_j(t)}+{\bbox{\mu}}_i(t),
     \label{eff1}
\end{equation}
  where the new Langevin noise ${\bbox{\mu}}$ is Gaussian with
  variance
\begin{eqnarray}
  \Bigl\langle{\bbox{\mu}}_i(t){\bbox{\mu}}_j(t')\Bigr\rangle=&&
    {2\over\zeta_0}\delta(t-t')\delta_{ij}\bbox{1}+2g_0^2
    \delta(t-t'){{\bf T}}({{\bf r}}_i(t),{{\bf R}}_j(t))
    \nonumber\\
    &&+{B^2\over 2\zeta_0^2}\delta_{ij}\int d1
    d2 C(t,t',1,2)
    {\partial U({{\bf r}}_i(t)-1)\over\partial{{\bf r}}_i(t)}
    {\partial U({{\bf r}}_i(t')-2)\over\partial{{\bf r}}_i(t')}
\label{re-noise}
\end{eqnarray}
   The effective Hamiltonian $H_{eff}$ becomes
\begin{eqnarray}
 H_{eff}=H_0+&&B_0\sum_i\int d1 {\bar\rho}({{\bf r}},t)
		U({{\bf r}}_i(t)-1)
	      \nonumber\\
	     &&-{B^2\over 2\zeta_0}\sum_i\int_\infty^{t^-}dt'\int
	       d1 d2 \bbox{G}(t,t',1,2)
	      {\bbox{\cdot}} {\partial U({{\bf r}}_i(t')-2)
	      \over\partial{{\bf r}}_i(t')}
	      U({{\bf r}}_i(t)-1)
\label{H-eff0}
\end{eqnarray}
   where $t^-\equiv t-\epsilon$, $\epsilon$ being an infinitesimal positive
   number.

   These equations have the form of the nonlinear generalized Langevin
   equation~\cite{Zwanzig:gle,Kawasaki:gle}. However, the reason for
   the emergence of this nonlinear generalized Langevin equation is
   different from that in Refs.\ \cite{Zwanzig:gle,Kawasaki:gle}. It
   is from the elimination of heterogeneity of the original system,
   rather than from the elimination of the degrees of freedom of
   surrounding "bath" molecules. Here the elimination of the
   surrounding "bath" molecules does not cause any memory effect
   because Markov approximation is used.

   In generalized hydrodynamics one can define frequency and wavelength
   dependent transport coefficients \cite{Boon}. Similarly, in Eq.\
   (\ref{H-eff0}), the term containing heterogeneity $B$ can be seen
   as giving a time-dependent excluded volume interaction. Alternatively,
   since this term is non-local in time, one can say that the heterogeneity
   introduces an excluded volume interaction along the world line, in
   addition to the usual excluded volume interaction $B_0$ along the
   chain contour.

   Inspecting these equations, we find that they are similar, albeit
   more complicated, to the effective kinetic equation obtained in
   spin glass dynamics \cite{Sompolinsky,Kirk:p-spin}. In these works
   a memory function also appears in mean-field limit. In principle,
   one can expand the monomer-monomer interaction $U$ in terms of
   dynamical variables ${{\bf r}}_i(t)$. This expansion will lead to
   a generalization of the $p$-spin-interaction spin-glass dynamics
   discussed in Ref.\ \cite{Kirk:p-spin}. However, since one will
   get infinite $p$-spin-interaction terms, this approach does not seem
   to be practically feasible.

   We can now take the known result for homopolymer as a reference.
   When we set the chain heterogeneity $B$ zero, Eq.\ (\ref{eff1}) is
   expected to reduce to the same equation obtained from eliminating
   the solvent velocity field in the Oono-Freed kinetic equation
   (See App.\ \ref{Kirkwood}),
\begin{equation}
  {\partial\over\partial t}{{\bf r}}_i(t)=
  -{1\over\zeta_0}{\partial H_E\over\partial {{\bf r}}_i(t)}
  -g_0^2\sum_j{{\bf T}}({{\bf r}}_i(t),{{\bf r}}_j(t))
   {\bbox{\cdot}}{\partial H_E\over\partial{{\bf r}}_j(t)}
  +{\bbox{\mu}}_i^E(t)
\end{equation}
\begin{equation}
  \Bigl\langle{\bbox{\mu}}_i^E(t){\bbox{\mu}}_j^E(t')\Bigr\rangle=
      {2\over\zeta_0}\delta(t-t')\delta_{ij}\bbox{1}
      + 2g_0^2\delta(t-t'){{\bf T}}({{\bf r}}_i(t),{{\bf r}}_j(t))
\end{equation}
   where $H_E$ is the Edwards Hamiltonian (\ref{HE}).
   This equation reveals that, the lowest order approximation of
   $\bar\rho({{\bf R}},t)$ must be the usual monomer density.
   Any higher order correction, after multiplied by $B_0$, must be
   first or higher order in $B_0$, $g_0^2/\eta_0$, or $B^2/\zeta_0$.
   To the order of our calculation, these corrections can be dropped.
   Therefore, we can replace $\bar\rho({{\bf R}},t)$ by its
   corresponding un-bracketed quantity:
\begin{equation}
  \bar\rho({{\bf R}},t)\equiv\sum_i\delta({{\bf r}}_i(t)-{{\bf R}})
\end{equation}

   Taking into account the symmetry of monomer indices, which
   should be restored when the self-consistent bracket is removed,
   the effective Hamiltonian (\ref{H-eff0}) becomes
\begin{eqnarray}
 H_{eff}=H_0+&&{B_0\over 2}\sum_{ij} U({{\bf r}}_i(t)-{{\bf r}}_j(t))
	       \nonumber\\
	     &&-{B^2\over 2\zeta_0}\sum_i\int_\infty^{t^-}dt'\int
	       d1 d2 \bbox{G}(t,t',1,2)
	      {\bbox{\cdot}} {\partial U({{\bf r}}_i(t')-2)\over
	      \partial{{\bf r}}_i(t')} U({{\bf r}}_i(t)-1)
\label{H-eff1}
\end{eqnarray}

   Since there is no corresponding known result for $B_0=0$ and $B\neq0$
   (a heteropolymer chain in theta solution), we cannot find a lowest
   order approximation for $\bbox{G}$ using the same method.

   The quantity $\bbox{G}(t,t',{{\bf R}}_1,{{\bf R}}_2)$
   in Eq.\ (\ref{H-eff1}) gives the response
   field of the chain at time $t$, when the chain conformation is
   $\delta({{\bf r}}_i(t)-{{\bf R}}_1)$, to a perturbation, introduced by
   ${{\bf {\hat{r}}}}_i(t')$ at time $t'$, when the chain conformation was
   $\delta({{\bf r}}_i(t')-{{\bf R}}_2)$. (The perturbation can be
   explicitly introduced in the kinetic equation, in the same way as
   the perturbative magnetic field used in \cite{Sompolinsky,Kirk:p-spin}.
   For simplicity it is not included here.) Causality again is respected
   and the integral of $t'$ in the effective Hamiltonian is limited
   by current time $t$.

   The quantity $C(t,t',{{\bf R}}_1,{{\bf R}}_2)$ in the correlation
   function of the renormalized noise ${\bbox{\mu}}_i$, Eq.\
   (\ref{re-noise}), gives the correlation function of chain
   conformations, between conformation
   $\delta({{\bf r}}_i(t)-{{\bf R}}_1)$ at time $t$ and
   $\delta({{\bf r}}_i(t')-{{\bf R}}_2)$ at time $t'$. This is a
   natural dynamical generalization of the static order parameter
   proposed in Ref.\ \cite{bio}.

   Note that the effective Hamiltonian is a time-dependent Hamiltonian.
   The memory function $\bbox{G}(t,t')$ makes the chain conformation at
   time $t$ depend on all the past chain conformations. This memory
   term makes the kinetic equation non-Markovian, and writing a
   Fokker-Planck equation for it impossible. It is well known that a
   non-Markovian process can be made Markovian by introducing new
   degrees of freedom (e.g. the exmaple in Ref.\ \cite{Wang}). However,
   in our case, this would be equivalent to going back to the original
   Oono-Freed kinetic equations where the process was Markovian,
   because what has been done is exactly elimination of some degrees
   of freedom (heterogeneity and solvent) from the original Markovian
   process. Of course, a Fokker-Planck-like equation can be written for
   Eqs.\ (\ref{eff1}) and (\ref{H-eff1}), but this would not give a
   closed equation. Instead, it gives an equation with infinite terms
   (e.g. a Kramers-Moyal expansion) \cite{Hanggi}.

   On the other hand, a Fokker-Planck equation with memory effect
   has been developed by Zwanzig \cite{Zwanzig}. It is also shown
   in Ref.\ \cite{Zwanzig} that, after certain
   approximations, the memory Fokker-Planck equation can be transformed into
   a generalized Langevin equation. This seems to allow us to transform
   the generalized Langevin equation obtained here back to a memory
   Fokker-Planck equation. However, it is not clear how to perform this
   backward transformation.

   These considerations make it clear that trying to write down a
   generalized (memory) Kirkwood diffusion equation for the mean field
   equation of motion Eq.\ (\ref{eff1}) would be very difficult, if
   not impossible. Fortunately, although the relaxation spectrum is
   calculated in Ref.\ \cite{spec} by perturbatively solving the
   Kirkwood diffusion equation, one can also derive the relaxation
   times directly from the kinetic equation. As it is clear from
   the above analysis, we should not try to find the corresponding
   generalized Kirkwood diffusion equation but rather derive the
   relaxation spectrum directly from the kinetic equation. This
   is done in the next section.

\section{Mode relaxation spectrum}
\subsection{Transition to continuous model}
\label{mode}

   For convenience, we now rewrite the effective equation of motion
   in terms of continuous chain notation. We will use the standard
   notation \cite{Oono:review,Freed:book}: $\tau$ being the contour
   length measured along the chain; ${{\bf c}}(\tau,t)$ the position
   of the point at $\tau$. A short distance cut-off $a$ will be
   implicitly assumed in the following equations so that
   $\mid\tau-\tau'\mid\, > a$ for all double integrals over $\tau$
   and $\tau'$. The upper limit of contour length can be $N-1$,
   according to Eq.\ (\ref{Hamiltonian}), or $N$, because $N$ has
   been taken to be very large in mean field approximation. We will
   use $N$ as the upper limit of contour length so that one can
   readily compare our result to the result of Ref.\ \cite{spec}.

   The effective equation of motion in continuous chain notation
   is
\begin{equation}
  {\partial\over\partial t}{{\bf c}}(\tau,t)=
    -{1\over\zeta_0}{\delta H_{eff}\over\delta{{\bf c}}(\tau,t)}
    -g_0^2\int d\tau'{{\bf T}}({{\bf c}}(\tau,t)-{{\bf c}}(\tau',t))
    {\bbox{\cdot}}{\delta H_{eff}\over\delta{{\bf c}}(\tau',t)}
    +{\bbox{\mu}}(\tau,t), \label{cont eq}
\end{equation}
   where
\begin{eqnarray}
  \Bigl\langle{\bbox{\mu}}(\tau,t){\bbox{\mu}}(\tau',t')\Bigr\rangle=
   &&{2\over\zeta_0}\bbox{1}\delta(t-t')\delta(\tau-\tau')
     +2g_0^2\delta(t-t'){{\bf T}}({{\bf c}}(\tau,t)-{{\bf c}}(\tau',t'))
    \nonumber\\
   &&+{B^2\over 2\zeta_0^2}\delta(\tau-\tau')\int d1 d2 C(t,t',1,2)
     {\delta U({{\bf c}}(\tau,t)-1)\over{{\bf c}}(\tau,t)}
     {\delta U({{\bf c}}(\tau',t')-2)\over{{\bf c}}(\tau',t')}
\end{eqnarray}
\begin{equation}
   H_{eff}=H_E-{B^2\over 2\zeta_0}\int d\tau\int_{-\infty}^{t^-} dt'
	   \int d1 d2 \bbox{G}(t,t',1,2){\bbox{\cdot}}
	   {\delta U({{\bf c}}(\tau,t')-2)\over\delta{{\bf c}}(\tau,t')}
	   U({{\bf c}}(\tau,t)-1)
\label{cont-H-eff}
\end{equation}
\begin{equation}
   H_E={1\over 2}\int d\tau \mid{\partial\over\partial\tau}
	 {{\bf c}}(\tau,t)\mid^2 + {B_0\over 2}\int d\tau d\tau'
	   U({{\bf c}}(\tau,t)-{{\bf c}}(\tau',t))
\end{equation}
\begin{equation}
   \bbox{G}(t,t',1,2)=\int d\tau \Bigl\langle\delta({{\bf c}}(\tau,t)-1)
		i{{\bf {\hat{c}}}}(\tau,t')\delta({{\bf c}}(\tau',t)-2)
		\Bigr\rangle
\end{equation}
\begin{equation}
   C(t,t',1,2)=\int d\tau \Bigl\langle\delta({{\bf c}}(\tau,t)-1)
			     \delta({{\bf c}}(\tau',t)-2)\Bigr\rangle
\end{equation}
\begin{equation}
   {{\bf T}}({{\bf c}}(\tau,t)-{{\bf c}}(\tau',t)) = \int_{{\bf k}}
	    {1\over\eta_0 k^2}{{\bf P}}^T({{\bf {\hat{k}}}})
	    e^{i{{\bf k}}{\bbox{\cdot}}({{\bf c}}(\tau,t)-{{\bf c}}(\tau',t))}
\end{equation}

   From now on the potential $U({{\bf c}}(\tau,t)-{{\bf c}}(\tau',t))$
   will be taken as a hard-core repulsive interaction $\delta({{\bf c}}
   (\tau,t)-{{\bf c}}(\tau',t))$ so that we can compare our result with
   Ref.\ \cite{spec}. For a general potential the corresponding equations
   can be easily written down.

   Eq.\ (\ref{cont eq}) essentially describes a non-equilibrium
   process. In spin glass dynamics, as mentioned above, a similar
   equation can be obtained through the same mean field approximation
   \cite{Sompolinsky,Kirk:p-spin}. In Refs.\
   \cite{Sompolinsky,Kirk:p-spin}, it is assumed that the process
   is stationary. While the newly developed out-of-equilibrium
   dynamics \cite{out-of} is potentially able to address
   non-stationary dynamics, we will assume that in our case
   the process is stationary. This assumption gives
\begin{equation}
   \bbox{G}(t,t',1,2)=\bbox{G}(t-t',1,2)
\end{equation}
\begin{equation}
   C(t,t',1,2)=C(t-t',1,2)
\end{equation}

\subsection{Perturbation calculation}
\label{pert}

   The conventional method used to determine the spectrum of mode
   relaxation, especially the longest relaxation time, is to find
   the eigenvalues and eigenfunctions for the Fokker-Planck
   operator \cite{spec}. This approach is useful as long as a
   Fokker-Planck equation exists. However, as we have analyzed
   in the previous section, this is not the case here.

   Nevertheless, since what we want to find is the longest
   relaxation time, we can simply use perturbation method to
   find out how the kinetic coefficient is renormalized, with
   the same spirit of Refs.\ \cite{Mori,Ma}. In principle, it is
   possible to develop a diagrammatic method, similar to the
   diagrammatics developed in Ref.\ \cite{Stepanow}, to systematically
   calculate desired quantities. However, this would not make much
   sense because Eqs.\ (\ref{cont eq}) is valid only to the
   lowest nontrivial order. A straightforward perturbation
   calculation will be sufficient.

   Introducing Rouse coordinates ${\bbox{\xi}}(p,t)$,
\begin{equation}
  {\bbox{\xi}}(p,t)\equiv \sqrt{{2\over N}}\int d\tau {{\bf c}}(\tau,t)
     \cos{p\pi\tau\over N}
\end{equation}
   and its Fourier transform
\begin{equation}
  {\bbox{\xi}}(p,\omega)\equiv \int dt {\bbox{\xi}}(p,t)e^{i\omega t},
\end{equation}
   the equation of motion becomes
\begin{eqnarray}
   G_0^{-1}(p,\omega){\bbox{\xi}}(p,\omega)=
   &&-{B_0\over\zeta_0}\sqrt{{2\over N}}\int d\tau d\tau' dt \int_{{\bf k}}
     e^{i\omega t}i{{\bf k}}\cos{p\pi\tau\over N}
   \nonumber\\
   &&\qquad\times\exp\Bigl[i\sqrt{{2\over N}}
     {{\bf k}}{\bbox{\cdot}}\sum_{p'}\int_{\omega'}e^{-i\omega t'}
     {\bbox{\xi}}(p',\omega')Q_{p'}(\tau,\tau')\Bigr]
   \nonumber\\
   &&-g_0^2{2\over N}\int d\tau d\tau' dt \int_{{\bf k}}\int_\omega\sum_{p'}
     e^{i\omega t}({p'\pi\over N})^2{1\over\eta_0 k^2}
     \cos{p\pi\tau\over N}\cos{p'\pi\tau'\over N}
   \nonumber\\
   &&\qquad\times\exp\Bigl[i\sqrt{{2\over N}}{{\bf k}}{\bbox{\cdot}}
     \sum_{p''}\int_{\omega''}
     e^{-i\omega'' t}{\bbox{\xi}}(p'',\omega'')Q_{p''}(\tau,\tau')\Bigr]
     {{\bf P}}^T({{\bf {\hat{k}}}}){\bbox{\cdot}}{\bbox{\xi}}
     (p',\omega')e^{-i\omega' t}
   \nonumber\\
   &&+{B^2\over 2\zeta_0^2}\sqrt{{2\over N}}\int dt d\tau d1 d2
     \int_{-\infty}^{t^-}dt'\int_{{\bf k}}\int_{{{\bf k}}'}
     e^{i\omega t}\bbox{G}(t,t',1,2){\bbox{\cdot}} i{{\bf k}}'i{{\bf k}}
     \cos{p\pi\tau\over N}
   \nonumber\\
   &&\qquad\times e^{-i{{\bf k}}'{\bbox{\cdot}}{{\bf R}}_2}
     \exp\Bigl[i\sqrt{{2\over N}}{{\bf k}}'{\bbox{\cdot}}
     \sum_{p''}\int_{\omega''}{\bbox{\xi}}(p'',\omega'')e^{-i\omega''t'}
     \cos{p''\pi\tau\over N}\Bigr]
   \nonumber\\
   &&\qquad\times e^{-i{{\bf k}}{\bbox{\cdot}}{{\bf R}}_1}
     \exp\Bigl[i\sqrt{{2\over N}}{{\bf k}}{\bbox{\cdot}}
     \sum_{p'}\int_{\omega'}{\bbox{\xi}}(p',\omega')e^{-i\omega't}
     \cos{p'\pi\tau\over N}\Bigr]
   \nonumber\\
   &&+{\bbox{\mu}}(p,\omega),
\label{Rouse-mode-eq}
\end{eqnarray}
   and
\begin{eqnarray}
     \Bigl\langle{\bbox{\mu}}(p,\omega){\bbox{\mu}}(p',\omega')\Bigr\rangle=
   &&{2\over N}{2\over\zeta_0}\bbox{1}\int dt d\tau e^{i(\omega+\omega')t}
	 \cos{p\pi\tau\over N}\cos{p'\pi\tau\over N}
   \nonumber\\
   +&&{2\over N}2g_0^2\int dt d\tau d\tau' e^{i(\omega+\omega')t}
	 \cos{p\pi\tau\over N}\cos{p'\pi\tau'\over N}
	 {{\bf T}}({{\bf c}}(\tau,t)-{{\bf c}}(\tau',t))
   \nonumber\\
   +&&{2\over N}{B^2\over 2\zeta_0^2} \int dt d\tau dt'
	 e^{i\omega t} e^{i\omega't'}
	 \cos{p\pi\tau\over N}\cos{p'\pi\tau\over N}
	 \int d1 d2 C(t,t',1,2)
   \nonumber\\
    &&\qquad\times
	    {\delta\delta({{\bf c}}(\tau,t)-1)\over\delta{{\bf c}}(\tau,t)}
	    {\delta\delta({{\bf c}}(\tau,t')-2)\over\delta{{\bf c}}(\tau,t')}
\label{mumu-full}
\end{eqnarray}
   where we have defined
\begin{equation}
   \int_\omega\equiv\int{d\omega\over 2\pi},
\end{equation}
\begin{equation}
   G_0^{-1}(p,\omega)\equiv -i\omega+ \Lambda_p,\qquad\qquad
   \Lambda_p\equiv {1\over\zeta_0}({p\pi\over N})^2,
\end{equation}
\begin{equation}
   Q_p(\tau,\tau')\equiv\cos{p\pi\tau\over N}-\cos{p\pi\tau'\over N}.
\end{equation}

   Since we need to carry out calculations only to the lowest
   nontrivial order, we can iterate this equation once and drop
   all the higher order terms. The detailed calculation can be
   found in App.\ \ref{1st}. It is shown in App.\ \ref{1st} that
   the relaxation spectrum $\Lambda_p$ is renormalized by three
   terms at the lowest nontrivial order,
\begin{eqnarray}
  &&
  \delta\Lambda_p^{(1)}=-{B_0\over\zeta_0}{2\over N}\int d\tau d\tau'
       \int_{{\bf k}} {k^2\over d}\cos{p\pi\tau\over N}Q_p(\tau,\tau')
       e^{-k^2B_N(\tau,\tau')}, \label{lambda1}
  \\
  &&
  \delta\Lambda_p^{(2)}=-{g_0^2\over\eta_0}{2\over N}({p\pi\over N})^2
       \int d\tau d\tau' \int_{{\bf k}} {1-1/d\over k^2}
       \cos{p\pi\tau\over N}\cos{p\pi\tau'\over N}
       e^{-k^2B_N(\tau,\tau')}, \label{lambda2}
  \\
  &&
  \delta\Lambda_p^{(3)}=-{B^2\over 2\zeta_0^2}{2\over N}\int d\tau
       \int_{{\bf k}}\cos{p\pi\tau\over N}\cos{p\pi\tau'\over N}
	i\Bigl[\bbox{\tilde G}(\tau,0,{{\bf k}})-
	  \bbox{\tilde G}(\tau,-\omega,{{\bf k}})\Bigr]
       {\bbox{\cdot}}{{\bf k}}{k^2\over d}, \label{lambda3}
\end{eqnarray}

   Compared to the result obtained in Ref.\ \cite{spec}, Eqs.\
   (\ref{lambda1}) and (\ref{lambda2}), renormalizations due to
   excluded volume interaction and hydrodynamic interaction respectively,
   coincide with the results reported there. The additional
   renormalization due to the chain heterogeneity $B$ is given by
   $\delta\Lambda_p^{(3)}$. It is found in Ref.\ \cite{spec} that
   the renormalizations from excluded volume and hydrodynamic
   effects are independent of each other, at the lowest nontrivial order.
   Here we see that the addition of heterogeneity preserves this
   independence: All three effects are not coupled.

   It has been shown that in spin glasses there exists a critical
   temperature below which the	replica symmetric solution is not
   stable \cite{spin}. Below this transition temperature, there are
   infinite ground states so that the system displays non-ergodic
   behavior. This non-ergodic behavior is responsible for many
   characteristics of the spin glasses below the transition temperature
   \cite{spin}. It is proposed in Ref.\ \cite{Sompolinsky} that,
   in an infinite system, the fluctuation-dissipation theorem (FDT)
   should be violated below the spin-glass transition temperature
   $T_c$. Mathematically, the violation comes from the extra
   time-persistent term in the correlation and response functions.
   Physically, the breaking of FDT is caused by the non-ergodic
   behavior of the system.
   The response function contains not only the usual response obeying
   the FDT but also a time-persistent part corresponding to the crossing of
   the energy barriers between the ground states.
   Ref.\ \cite{Sompolinsky} also shows that the breaking of FDT is
   needed in order to make the static mean-field solution
   of the Sherrington-Kirkpatrick (SK) model stable.

   It is clear that this dynamical consequence (i.e. breaking of FDT)
   is closely related to the static property (many ground states)
   of the SK model of spin glasses. Ref.\ \cite{bio} shows that a
   freezing transition can occur in heteropolymers at which only
   few conformations (states) dominate in the partition function.
   One might expect some dynamic consequences, and in fact it was
   suggested indeed \cite{bwkin} that static freezing in heteropolymers
   is accompanied by that dynamic freezing. This assertion was made
   based on the phenomenological analysis which extends the REM
   to kinetics. The model used in \cite{bwkin} employs an unphysical
   assumption that two states connected by one kinetic step have
   statistically independent energies; it is this assumption which
   leads to the conclusion about the glass transition in this
   system \cite{Gutin}. Our microscopic analysis, Eq.\ (\ref{lambda3})
   shows that, within the order of our calculations whether the
   FDT is broken or not, dynamically, there is no effect at all.
   This is because the $\bbox{\tilde{G}}(\tau,0,{{\bf k}})$ term
   exactly cancels the time-persistent part of the response field,
   if the temperature is below the transition temperature. Since the
   time-persistent part of response function is directly related to
   the degeneracy of the ground states, this result implies that this
   degeneracy is dynamically unobservable, at least to the lowest
   nontrivial order. This, in turn may mean that barriers between
   low-energy states are nonextensive (in $N$) in the studied model
of heteroplymer.

   One may argue that the violation of FDT happens only when the system
   is infinitely large, $N=\infty$, and, since only large $N$ is imposed
   in the derivation of the effective equation of motion, breaking of
   FDT does not happen and the above argument is not quite right.
   Nevertheless, what really matters is that the term
   $\bbox{\tilde{G}} (\tau,0,{{\bf k}})
   -\bbox{\tilde{G}}(\tau,-\omega,{{\bf k}})$ makes all static effects
   at zero frequencies dynamically null, and corresponding static
   freezing phenomena may not have its have its dynamic counterpart
   in this system.

   Carefully checking the calculation that leads to
   the term $\bbox{\tilde{G}} (\tau,0,{{\bf k}})
   -\bbox{\tilde{G}}(\tau,-\omega,{{\bf k}})$ in Eq.\ (\ref{lambda3}),
   we see that it is rooted in the two exponential
   functions in the $B$-dependent term in Eq.\ (\ref{Rouse-mode-eq}).
   This makes it clear that the chain flexibility is
   responsible for the dynamically null result. Physically, this is
   a plausible result. The quenched randomness in
   Eq.\ (\ref{Hamiltonian}) is quenched along the chain, but not
   in the space in which the heteropolymer is embedded. As we do
   not expect a dilute liquid mixture of magnetic and non-magnetic
   particles displaying any ``frozen in'' behavior observed in its
   solid mixture counterpart, e.g. spin glasses, we do not expect
   a full spin-glass-like behavior in heteropolymers because the
   randomness is not fully quenched in space. We probably can say that the
   correspondence between spin glasses and heteropolymers is statically
   more complete than dynamically. The analogy between spin glasses
   and biopolymers therefore has to be used with some care, especially
   in building {\em ad hoc} phenomenological models not supported
   by microscopic analysis.

\section{On the analogy between spin glass and heteropolymer}
\label{analogy}

   It is well known that there exists an elegant correspondence between
   magnetic system and self-avoiding random walk, e.g. a polymer chain
   with excluded volume effect. This correspondence
   can be shown perturbatively and analytically
   \cite{Cloizeaux,deGennes,Emery-Fisher}.
   From the point of view of this complete correspondence, it seems that
   the analogy between spin glass and heteropolymer should be also
   complete. However, according the result obtain in the prvious section,
   one has to be careful when using this analogy. Therefore, it is
   necessary to think more about this analogy in the context
   of the correspondence.

   There are different ways to establish the correspondence
   between magnetic system and self-avoiding random walk \cite{Cloizeaux}.
   Here we will follow the analytical proof given in
   Ref.\ \cite{Emery-Fisher}. Using the result obtained in
   Ref.\ \cite{Emery-Fisher} we know that
   the model considered here corresponds to an Ising model
   with non-random nearest neighbor interaction and random four-spin
   coupling. More precisely, Eqs.\ (\ref{Hamiltonian}) can be
   obtained from the $n=0$ limit of the following soft-spin Hamiltonian,
\begin{equation}
   H=\sum_\alpha^n\sum_{ij}^N J_{ij}\sigma_{i\alpha}\sigma_{j\alpha}
   +\sum_i^N\Bigl[r\sum_\alpha^n \sigma_{i\alpha}^2 + {B_{ij}\over 2}
   (\sum_\alpha^n \sigma_{i\alpha}^2)^2\Bigr]
\label{emery}
\end{equation}
   where $r$ serves as a variable for Laplace transform (in getting
   the partition function for the walk) and its value needs not to
   be specified in our discussion. Further detail
   of this correspondence can be found in \cite{Emery-Fisher}.

   Comparing this Hamiltonian with the one considered in spin
   glass dynamics \cite{Sompolinsky}, in terms of soft-spin,
\begin{equation}
   H={1\over 2}\sum_{\langle ij \rangle}(r_0\delta_{ij}-2\beta J_{ij})
     \sigma_i\sigma_j + u\sum_i\sigma_i^4-\sum_i h_i\sigma_i
\label{somp}
\end{equation}
   we see that the main difference is where the randomness enters.
   Eq.\ (\ref{somp}) has the quenched randomness $J_{ij}$ being
   a two-spin coupling, while the random coupling $B_{ij}$
   in Eq.\ (\ref{emery}) is in 4-spin term. Note that the Hamiltonian
   (\ref{emery}) is also different from the $p$-spin-interaction
   spin glass studied in Ref.\ \cite{Kirk:p-spin},
\begin{equation}
   H=-\sum_{i_1<i_2\cdots<i_p}J_{i_1\cdots i_p}\sigma_{i_1}\cdots\sigma_{i_p}
     -\sum_{i=1}^N h_i\sigma_i
\label{kirk1}
\end{equation}
   (or, in terms of soft-spin,
\begin{equation}
   H=\sum_i\bigl[{r_0\over 2}\alpha_i^2+u\sigma_i^4\bigr]-\beta\sum
     J_{i_1\cdots i_p}\sigma_{i_1}\cdots\sigma_{i_p} - \beta\sum_{i=1}^N
     h_i\sigma_i
\label{kirk2}
\end{equation}
   ). Although Eq.\ (\ref{kirk2}) appears to be very similar to
   Eq.\ (\ref{emery}), the additional two-spin coupling $J_{ij}$ and
   the $O(n)$-symmetry preserving term $B_{ij}$ in Eq.\ (\ref{emery})
   make the story entirely different. It is unclear, and
   seems unlikely, that the (qualitative) dynamic features obtained in
   Refs.\ \cite{Sompolinsky,Kirk:p-spin} have their counterparts in
   the dynamics of Eq.\ (\ref{emery}). In fact, we are not aware of any
   study of Eq.\ (\ref{emery}) itself as a possible spin glass model.
   Perhaps Eq.\ (\ref{emery}) being a spin glass model has not
   been studied, either statically or dynamically, before.

   (Perhaps the $^3$He-$^4$He-aerogel system mentioned in Introduction
   is more relevant to the heteropolymer solution than spin glass system.
   We argued in the previous section that the randomness in heteropolymer
   solution is not completely quenched in space. In $^3$He-$^4$He-aerogel
   system, according to the picture proposed by Chan and coworkers
   \cite{helium}, at low $^4$He concentration, $^4$He atoms can be seen as
   ``partially quenched'': Some are bound onto the aerogel, being
   completely quenched, while some are in freely moving superfluid
   phase. This interplay of quenched randomness and annealed
   randomness has been theoretically studied by Maritan {\it et. al.}
   \cite{Maritan}, using the so-called Blume-Emery-Griffiths
   (BEG) Hamiltonian \cite{BEG}. Interestingly, the similarity between
   the BEG Hamiltonian \cite{Maritan,BEG} and Eq.\ (\ref{emery})
   seems higher than Eqs.\ (\ref{somp}) and (\ref{kirk2}).)

   The above observation is only at the level of the Hamiltonian
   used in each model. In fact, more significant difference at the
   level of kinetic equation exists. In the Oono-Freed kinetic
   equation (\ref{O-F eq}) the dynamical variables are monomer
   positions $\{{{\bf r}}_i\}$ and solvent velocity field ${{\bf u}}$.
   On the other hand, the dynamical variables used in Refs.\
   \cite{Sompolinsky,Kirk:p-spin} are spin components $\{\sigma_i\}$.
   When $n\to 0$, spins $\{\sigma_i\}$ do not become monomer positions
   $\{{{\bf r}}_i\}$ and solvent velocity field ${{\bf u}}$. Although
   the effective equation of motion does share some common features,
   e.g., memory effect, renormalized noise, etc., with the ones found
   in spin glass study, these are the features common to the machinery,
   functional integral formalism plus mean field approximation.
   Having these common features still allows entirely different physics.
   As we argued at the end of Sec.\ \ref{mode}, our system is somewhat
   intermediate between the solid and molten phases of dilute magnets,
   and is perhaps more similar to $^3$He-$^4$He mixture in porous media
   than spin glasses.

\section{Conclusion}
\label{disc}

   The main result of this paper is the effective equation of motion
   Eqs.\ (\ref{eff0}) and (\ref{eff1}) and the renormalized relaxation
   spectrum.

   We first ask how adaptive our model is. In polymer dynamics much
   work has been done on homopolymers. There also have been some work
   on the dynamics of copolymers. As a first approximation, if the
   heterogeneity of the chain is not important, homopolymer is a
   reasonable model. It is seen in the above calculation that the
   lowest order effect of the heterogeneity parameter $B$ is
   proportional to $B^2$. This also indicates that this approximation
   makes sense. It is also known that the natural-occurred heteropolymers
   such as proteins are composed of at most 20 types of monomers
   (amino acids). In this sense the approximation of independent
   interaction energies $B_{ij}$ employed in this paper may be
   reasonable: It was shown \cite{bio} that the greater the number of
   monomer types is, the better the approximation of independent
   $B_{ij}$ is.

   Furthermore, in a recent simulation \cite{Dawson:kin}, it is shown
   that in the early stage of the collapse of homopolymer, from coil
   state to globule state, some clusters form along the chain. One way
   to see this process is to view the chain, with clusters of various
   sizes, as an effective heteropolymer chain formed by monomers and
   clusters. This is plausible if one uses a coarse-grained viewpoint.
   One may argue that these clusters are not stable; they constantly form
   and annihilate. However, on average and in a coarse-grained sense,
   this constant birth-and-death of clusters probably can be ignored
   because the clusters seldom grow larger than certain scale, as
   the simulation shows. We may take the upper length scale of these
   clusters as a basic length scale in the model, and treat the clusters
   as heterogeneous monomers. These clusters may have different sizes
   and masses, which are also allowed in our model because sizes and
   masses of monomers do not appear in the Oono-Freed kinetic equation.
   In this sense, our model can also serve as a model for a homopolymer
   chain in poor solvent during its early stage of collapse.

   Ideally, similar to Ref.\ \cite{spec}, the next calculation would
   be applying renormalization group technique to find the scaling
   form of the relaxation spectrum. This does not seem promising,
   however. To perform this calculation, we need an ansatz for the
   $\omega$-dependence of $\bbox{G}(\omega,{{\bf k}})$, analogous to
   the one used in spin glass dynamics \cite{Sompolinsky}, as well
   as an ansatz for its ${{\bf k}}$-dependence. Nevertheless, we do
   not expect that the renormalization group calculation, even if
   it is possible, will change the conclusion, regarding the dynamically
   null effect of the time-persistent part of response function,
   reached in Sec.\ \ref{pert}.

   The most important development of the present model would be to
   incorporate chain compactisation, i.e. consider the dynamics of the
   heteropolymer globule. Another relevant and important aspect of
   polymer dynamics which may be studied along these lines is kinetics
   of coil-globule transitions \cite{ganazzoli} which may be relevant
   also for protein folding.

\acknowledgments

   One of us (J.-R.R.) would like to thank Prof. Y. Oono on
   discussion regarding App.\ \ref{Kirkwood} and Ref.\ \cite{Oono-email}.

\appendix
\section{Derivation of Kirkwood diffusion equation}
\label{Kirkwood}

  We show here that the Kirkwood diffusion equation can be derived
  directly without the projection technique used in Refs.\
  \cite{Oono:review,Lee-Baldwin-Oono}

  If the Hamiltonian is the Edwards Hamiltonian, the equation of
  motion derived from Eqs.\ (\ref{O-F eq})
  will be (cf. Eq.\ (\ref{eq of motion})
\begin{equation}
  {\partial\over\partial t}{{\bf r}}_i(t)=-{1\over\zeta_0}{\partial H_E\over
  \partial{{\bf r}}_i(t)}+g_0{{\bf u}}_0^E({{\bf r}}_i(t),t)+
  {{\bbox{\theta}}_0}_i(t)+g_0{{\bf u}}_R^E({{\bf r}}_i(t),t),
		      \label{eq of motion-Edwards}
\end{equation}
\begin{equation}
  {{\bf u}}_0^E({{\bf r}},t)=
     \int_{{\bf k}} e^{i{{\bf k}}{\bbox{\cdot}}{{\bf r}}}\int_{-\infty}^t dt'
      e^{-\nu_0k^2(t-t')}{{\bf P}}^T({{\bf {\hat{k}}}})
      {\bbox{\cdot}}\Bigl[-{g_0\over\rho_0}
      \sum_{i=0}^{N-1}{\partial H_E\over\partial{{\bf r}}_i(t')}
      e^{-i{{\bf k}}{\bbox{\cdot}}{{\bf r}}_i(t')}\Bigr],
\end{equation}
\begin{equation}
  {{\bf u}}_R^E({{\bf r}},t)=
      \int_{{\bf k}} e^{i{{\bf k}}{\bbox{\cdot}}{{\bf r}}}\int_{-\infty}^t dt'
      e^{-\nu_0k^2(t-t')}{{\bf P}}^T({{\bf {\hat{k}}}})
      {\bbox{\cdot}}{\bbox{\xi}}_0({{\bf k}},t')
\end{equation}
   When Markov approximation is applicable,
\begin{equation}
   e^{-\nu_0k^2(t-t')}\to {2\over\nu_0 k^2}\delta(t-t')
\end{equation}
   the equation of motion becomes
\begin{equation}
  {\partial\over\partial t}{{\bf r}}_i(t)=-{1\over\zeta_0}{\partial H_E\over
  \partial{{\bf r}}_i(t)}-g_0^2\sum_j{{\bf T}}({{\bf r}}_i(t),{{\bf r}}_j(t))
  {\partial H_E\over\partial {{\bf r}}_j(t)}+{{\bbox{\theta}}_0}_i(t)
  +g_0\int_{{\bf k}} e^{i{{\bf k}}{\bbox{\cdot}}{{\bf r}}_i(t)}{1\over\eta_0
   k^2} {{\bf P}}^T({{\bf {\hat{k}}}}){\bbox{\cdot}}{\bbox{\xi}}({{\bf k}},t),
			 \label{Ed-Markov}
\end{equation}

   Eq.\ (\ref{Ed-Markov}) is a Langevin equation with noise
   ${\bbox{\theta}}^E({{\bf r}}_i(t),t)\equiv{{\bbox{\theta}}_0}_i(t)
   +g_0{{\bf u}}_R^E({{\bf r}}_i(t),t)$
   satisfying
\begin{equation}
  \bigl\langle{\bbox{\theta}}^E({{\bf r}}_i(t),t){\bbox{\theta}}^E
   ({{\bf r}}_j(t'),t')\bigr\rangle
  ={2\zeta_0}\delta(t-t')\delta_{ij}\bbox{1}
  +2g_0^2{{\bf T}}({{\bf r}}_i(t),{{\bf r}}_j(t'))\delta(t-t')
\end{equation}
   The Fokker-Planck equation corresponding to this Langevin equation
   is
\begin{equation}
  {\partial\over\partial t}P(\{{{\bf r}}_i\},t)=
    \sum_i\sum_j{\partial\over\partial{{\bf r}}_i}{\bbox{\cdot}}
    \Bigl[\zeta^{-1}\bbox{1}\delta_{ij}+g_0^2{{\bf T}}
    ({{\bf r}}_i,{{\bf r}}_j)\Bigr]
    {\bbox{\cdot}} \Bigl[{\partial\over\partial{{\bf r}}_j}+
    {\partial H_E\over\partial{{\bf r}}_j}\Bigr]
    P(\{{{\bf r}}_i\},t)
\label{app-kirk}
\end{equation}
  where $P(\{{{\bf r}}_i\},t)$ is the probability distribution
  function of chain conformation $\{{{\bf r}}_i\}$).
  Eq.\ (\ref{app-kirk}) is exactly the Kirkwood diffusion equation.

\section{Derivation of the effective Lagrangian}
\label{app:mf}

   Here we detail the calculation.

\subsection{$\sum_{i\neq j}{\cal O}_{ij}^2$ term}

   Define
\begin{eqnarray}
  &&\bar{A}(t,t',{{\bf R}},{{\bf R}}')\equiv\sum_i a_i(t)a_i(t')
      \delta({{\bf r}}_i(t)-{{\bf R}})\delta({{\bf r}}_i(t')-{{\bf R}}')
    \nonumber\\
  &&\bar{B}(t,t',{{\bf R}},{{\bf R}}')\equiv\sum_i
      \delta({{\bf r}}_i(t)-{{\bf R}})\delta({{\bf r}}_i(t')-{{\bf R}}')
    \nonumber\\
  &&\bar{C}(t,t',{{\bf R}},{{\bf R}}')\equiv\sum_i a_i(t)
      \delta({{\bf r}}_i(t)-{{\bf R}})\delta({{\bf r}}_i(t')-{{\bf R}}')
    \nonumber\\
  &&\bar{D}(t,t',{{\bf R}},{{\bf R}}')\equiv\sum_i a_i(t')
      \delta({{\bf r}}_i(t)-{{\bf R}})\delta({{\bf r}}_i(t')-{{\bf R}}')
    \nonumber\\
\end{eqnarray}
  Note that
\begin{eqnarray}
  &&\bar{A}(t,t',{{\bf R}},{{\bf R}}')=
    \bar{A}(t',t,{{\bf R}}',{{\bf R}})  \nonumber\\
  &&\bar{B}(t,t',{{\bf R}},{{\bf R}}')=
    \bar{B}(t',t,{{\bf R}}',{{\bf R}})  \nonumber\\
  &&\bar{C}(t,t',{{\bf R}},{{\bf R}}')=
    \bar{D}(t',t,{{\bf R}}',{{\bf R}})
\end{eqnarray}
  and they are all of order $O(N^2)$. Then,
\begin{eqnarray}
   \sum_{i\neq j}{\cal O}_{ij}^2={1\over 2}\int dt dt'
	   d1 d2 d3 d4
       \times \bigl[ && \bar{A}(t,t',1,3)
		   \bar{B}(t,t',2,4)
	      +    \bar{A}(t,t',2,3)
		   \bar{B}(t,t',1,4)  \nonumber\\
	      + && \bar{A}(t,t',1,4)
		   \bar{B}(t,t',2,3)
	      +    \bar{A}(t,t',2,4)
		   \bar{B}(t,t',1,3)  \nonumber\\
	      + && \bar{C}(t,t',1,3)
		   \bar{D}(t,t',2,4)
	      +    \bar{C}(t,t',2,3)
		   \bar{D}(t,t',1,4)  \nonumber\\
	      + && \bar{C}(t,t',1,4)
		   \bar{D}(t,t',2,3)
	      +    \bar{C}(t,t',2,4)
		   \bar{D}(t,t',1,3)  \bigr]  \nonumber\\
	      \times &&
		U(1-2)U(3-4) + O(N)
\end{eqnarray}

   Let column matrix $\psi(t,t',1,2,3,4)$ and $16\times 16$ matrix $S_1$ be
\begin{equation}
  \psi_A (t,t',1,2,3,4)\equiv\pmatrix{
    \bar{A}(t,t',1,3)\cr    \bar{A}(t,t',2,3)\cr
    \bar{A}(t,t',1,4)\cr    \bar{A}(t,t',2,4)\cr }, \qquad
  \psi_B (t,t',1,2,3,4)\equiv\pmatrix{
    \bar{B}(t,t',2,4)\cr    \bar{B}(t,t',1,4)\cr
    \bar{B}(t,t',2,3)\cr    \bar{B}(t,t',1,3)\cr },
\end{equation}
\begin{equation}
  \psi_C (t,t',1,2,3,4)\equiv\pmatrix{
    \bar{C}(t,t',1,3)\cr    \bar{C}(t,t',2,3)\cr
    \bar{C}(t,t',1,4)\cr    \bar{C}(t,t',2,4)\cr }, \qquad
  \psi_D (t,t',1,2,3,4)\equiv\pmatrix{
    \bar{D}(t,t',2,4)\cr    \bar{D}(t,t',1,4)\cr
    \bar{D}(t,t',2,3)\cr    \bar{D}(t,t',1,3)\cr }
\end{equation}
\begin{equation}
  \psi (t,t',1,2,3,4)\equiv\pmatrix{
    \psi_A(t,t',1,2,3,4)\cr    \psi_B(t,t',1,2,3,4)\cr
    \psi_C(t,t',1,2,3,4)\cr    \psi_D(t,t',1,2,3,4)\cr },\qquad
  S_1\equiv\pmatrix{0	& 1_4 &  0  &  0  \cr
	     1_4 &   0 &  0  &	0  \cr
	      0  &   0 &  0  & 1_4 \cr
	      0  &   0 & 1_4 &	0  \cr}
\end{equation}
   (where $1_4$ is a $4\times 4$ unit matrix), then
\begin{equation}
   \sum_{i\neq j}{\cal O}_{ij}^2={1\over 4}\int dt dt'
	   d1 d2 d3 d4
	   \psi^T(t,t',1,2,3,4) S_1
	     \psi(t,t',1,2,3,4)
		U(1-2)U(3-4) + O(N)
\end{equation}

\subsection{$\sum_{i\neq j}{\cal O}_{ij}$ term}

  The calculation for $\sum_{i\neq j}{\cal O}_{ij}$ can be performed
  in a similar way. However, because of the symmetry of $\bar{C}$
  and $\bar{D}$, there is an ambiguity in expressing
  $\sum_{i\neq j}{\cal O}_{ij}$ as a sum of products of $\bar{C}$
  and $\bar{B}$, or of $\bar{C}$ and $\bar{B}$, or a mixed type.
  This should not make the final result different, as we expect
  the same symmetry can be used to convert $\bar{C}$ and $\bar{D}$.

  If we express $\sum_{i\neq j}{\cal O}_{ij}$ in terms of products of
  $\bar{C}$ and $\bar{B}$, then
\begin{equation}
   \sum_{i\neq j}{\cal O}_{ij}={1\over 4}\int dt dt'
	   d1 d2 d3 d4
	   \psi^T(t,t',1,2,3,4) S_2
	     \psi(t,t',1,2,3,4)
		U(1-2)\bar{\delta}(t-t') + O(N)
\end{equation}
\begin{equation}
  S_2\equiv\pmatrix{0  &  0  &	0  &  0  \cr
	       0  &  0	& 1_4 &  0  \cr
	       0  & 1_4 &  0  &  0  \cr
	       0  &  0	&  0  &  0  \cr}
\end{equation}
   where the overbarred delta function $\bar{\delta}$ is used to
   keep track the fact that the integral over time variables,
   hence, the $\delta$-function, has to
   be done after the integral over space variables has been performed.

\subsection{Gaussian transform and mean-field approximation}

   Combining the results from the previous two sections,
   $\langle e^{L_R}\rangle_B$ in Eq.\ (\ref{Z after B}) can be written as
\begin{equation}
  \exp\Bigl\{\int dt dt' d1 d2 d3 d4
	    \bigl[{B^2\over 8\zeta_0^2}\psi^T S_1\psi U(1-2)U(3-4)
	    {B_0\over 4\zeta_0}  \psi^T S_2\psi U(1-2)\bar{\delta}(t-t')\bigr]
      +O(N)\Bigr\}
\label{all-L}
\end{equation}
   Introducing variables $Q_1(1,2,3,4) \cdots Q_{16}(1,2,3,4)$
   and performing a Gaussian transform, the generating functional
   $Z_{\theta_0\xi_0}$ becomes
\begin{equation}
   Z_{\theta_0\xi_0}=\int \{{\cal D}{{\bf r}}_i\}
	 \{{\cal D}{{\bf {\hat{r}}}}_i\}\{{\cal D}Q_i\}
	 \exp \bigl[-{1\over 4}{\cal Q}^T{\cal A}{\cal Q}
	       +{\cal Q}^T\psi + L_0 + O(N)\bigr]
\end{equation}
   where
\begin{equation}
  \{{\cal D}Q_i\} \equiv \prod_{i=1}^{16}{\cal D}Q_i
\end{equation}
   ${\cal Q}$ is the column matrix formed by $Q_1 \cdots Q_{16}$ and
   ${\cal A}$ is the coefficient of the quadratic term in Eq.\ (\ref{all-L}).

   Using steepest decent method the mean-field approximation gives
\begin{eqnarray*}
&&Q_1^0(t,t',1,2,3,4) ={B^2\over 4\zeta_0^2}U(1-2)U(3-4)
				  \bigl\langle\bar{B}(t,t',2,4)\bigr\rangle
     \nonumber\\
&&Q_2^0(t,t',1,2,3,4) ={B^2\over 4\zeta_0^2}U(1-2)U(3-4)
				  \bigl\langle\bar{B}(t,t',1,4)\bigr\rangle
     \nonumber\\
&&Q_3^0(t,t',1,2,3,4) ={B^2\over 4\zeta_0^2}U(1-2)U(3-4)
				  \bigl\langle\bar{B}(t,t',2,3)\bigr\rangle
     \nonumber\\
&&Q_4^0(t,t',1,2,3,4) ={B^2\over 4\zeta_0^2}U(1-2)U(3-4)
				  \bigl\langle\bar{B}(t,t',1,3)\bigr\rangle
     \nonumber\\
&&Q_5^0(t,t',1,2,3,4) ={B^2\over 4\zeta_0^2}U(1-2)U(3-4)
				  \bigl\langle\bar{A}(t,t',1,3)\bigr\rangle
		      +{B_0\over 2\zeta_0}U(1-2)\bar{\delta} (t-t')
				  \bigl\langle\bar{C}(t,t',1,3)\bigr\rangle
     \nonumber\\
&&Q_6^0(t,t',1,2,3,4) ={B^2\over 4\zeta_0^2}U(1-2)U(3-4)
				  \bigl\langle\bar{A}(t,t',2,3)\bigr\rangle
		      +{B_0\over 2\zeta_0}U(1-2)\bar{\delta} (t-t')
				  \bigl\langle\bar{C}(t,t',2,3)\bigr\rangle
     \nonumber\\
&&Q_7^0(t,t',1,2,3,4) ={B^2\over 4\zeta_0^2}U(1-2)U(3-4)
				  \bigl\langle\bar{A}(t,t',1,4)\bigr\rangle
		      +{B_0\over 2\zeta_0}U(1-2)\bar{\delta} (t-t')
				  \bigl\langle\bar{C}(t,t',1,4)\bigr\rangle
     \nonumber\\
&&Q_8^0(t,t',1,2,3,4) ={B^2\over 4\zeta_0^2}U(1-2)U(3-4)
				  \bigl\langle\bar{A}(t,t',2,4)\bigr\rangle
		      +{B_0\over 2\zeta_0}U(1-2)\bar{\delta} (t-t')
				  \bigl\langle\bar{C}(t,t',2,4)\bigr\rangle
     \nonumber\\
&&Q_9^0(t,t',1,2,3,4) ={B^2\over 4\zeta_0^2}U(1-2)U(3-4)
				  \bigl\langle\bar{D}(t,t',2,4)\bigr\rangle
		      +{B_0\over 2\zeta_0}U(1-2)\bar{\delta} (t-t')
				  \bigl\langle\bar{B}(t,t',2,4)\bigr\rangle
     \nonumber\\
&&Q_{10}^0(t,t',1,2,3,4) ={B^2\over 4\zeta_0^2}U(1-2)U(3-4)
				  \bigl\langle\bar{D}(t,t',1,4)\bigr\rangle
			 +{B_0\over 2\zeta_0}U(1-2)\bar{\delta} (t-t')
				  \bigl\langle\bar{B}(t,t',1,4)\bigr\rangle
     \nonumber\\
&&Q_{11}^0(t,t',1,2,3,4) ={B^2\over 4\zeta_0^2}U(1-2)U(3-4)
				  \bigl\langle\bar{D}(t,t',2,3)\rangle
			 +{B_0\over 2\zeta_0}U(1-2)\bar{\delta} (t-t')
				  \bigl\langle\bar{B}(t,t',2,3)\bigr\rangle
     \nonumber\\
&&Q_{12}^0(t,t',1,2,3,4) ={B^2\over 4\zeta_0^2}U(1-2)U(3-4)
				  \bigl\langle\bar{D}(t,t',1,3)\rangle
			 +{B_0\over 2\zeta_0}U(1-2)\bar{\delta} (t-t')
				  \bigl\langle\bar{B}(t,t',1,3)\bigr\rangle
     \nonumber\\
&&Q_{13}^0(t,t',1,2,3,4) ={B^2\over 4\zeta_0^2}U(1-2)U(3-4)
				  \bigl\langle\bar{C}(t,t',1,3)\bigr\rangle
     \nonumber\\
&&Q_{14}^0(t,t',1,2,3,4) ={B^2\over 4\zeta_0^2}U(1-2)U(3-4)
				  \bigl\langle\bar{C}(t,t',2,3)\bigr\rangle
     \nonumber\\
&&Q_{15}^0(t,t',1,2,3,4) ={B^2\over 4\zeta_0^2}U(1-2)U(3-4)
				  \bigl\langle\bar{C}(t,t',1,4)\bigr\rangle
     \nonumber\\
&&Q_{16}^0(t,t',1,2,3,4) ={B^2\over 4\zeta_0^2}U(1-2)U(3-4)
				  \bigl\langle\bar{C}(t,t',2,4)\bigr\rangle
     \nonumber\\
\end{eqnarray*}
   where the angular bracket means
\begin{equation}
  \langle\cdots\rangle={\int
       \{{\cal D}{{\bf r}}_i\} \{{\cal D}{{\bf {\hat{r}}}}_i\}(\cdots)
	     e^{L_0}e^{{\cal Q}^T\psi}
       \over \int
       \{{\cal D}{{\bf r}}_i\} \{{\cal D}{{\bf {\hat{r}}}}_i\}
	     e^{L_0}e^{{\cal Q}^T\psi} },
\label{average}
\end{equation}
   The mean-field solution for $Q_i^0$ are determined self-consistently
   from these equations.

   Therefore, the effective Lagrangian $L_e\equiv L_0+({\cal Q}^0)^T\psi$
   is
\begin{eqnarray}
  L_e=L_0&&+{B^2\over\zeta_0^2}\int dt dt'd1 d2 d3 d4 U(1-2)U(3-4)\bigl[
      \bigl\langle\bar{A}(t,t',1,3)\bigr\rangle
	     \bar{B}(t,t',2,4)  \nonumber\\
	&&\qquad\qquad+\bigl\langle\bar{B}(t,t',2,4)\bigr\rangle
	     \bar{A}(t,t',1,3) +
     2\bigl\langle\bar{C}(t,t',1,3)\bigr\rangle
	     \bar{D}(t,t',2,4)\bigr] \nonumber\\
	 &&+{2B_0\over\zeta_0}\int dt dt'd1 d2 d3 d4 \delta(t-t')
      U(1-2)\bigl[
      \bigl\langle\bar{C}(t,t',1,3)\bigr\rangle
	     \bar{B}(t,t',2,4)  \nonumber\\
	&&\qquad\qquad+\bigl\langle\bar{B}(t,t',2,4)\bigr\rangle
	     \bar{C}(t,t',1,3) \bigr],
\end{eqnarray}

\section{First order correction of relaxation spectrum}
\label{1st}

   The equation of motion in Sec.\ \ref{pert} has the following form
\begin{equation}
  G_0^{-1}(p,\omega){\bbox{\xi}}(p,\omega)={\bbox{\mu}}(p,\omega)
       +\bbox{F}[{\bbox{\xi}},p,\omega]
\end{equation}
   Substitute the zeroth order solution
\begin{equation}
  G_0^{-1}(p,\omega){\bbox{\xi}}^{(0)}(p,\omega)={\bbox{\mu}}(p,\omega)
\label{zeroth}
\end{equation}
   into it, the first order solution is
\begin{equation}
  {\bbox{\xi}}^{(1)}(p,\omega)=G_0(p,\omega){\bbox{\mu}}(p,\omega)
       +G_0(p,\omega)\bbox{F}[{\bbox{\xi}}^{(0)},p,\omega]
\end{equation}
   This solution contains terms to the lowest nontrivial order.
   Since ${\bbox{\xi}}$ enters in $\bbox{F}$ through exponential function,
   further iteration does not change the result to the lowest
   nontrivial order. Therefore, to the order we seek for, the solution
   is
\begin{equation}
  {\bbox{\xi}}(p,\omega)=G_0(p,\omega){\bbox{\mu}}(p,\omega)
       +G_0(p,\omega)\bbox{F}[{\bbox{\xi}}^{(0)},p,\omega]+\hbox{higher
       orders in}
\end{equation}

   To find the relaxation spectrum, we find the average (over
   ${\bbox{\mu}}$ of the following tensor product
\begin{eqnarray}
  {\bbox{\xi}}(p,\omega){\bbox{\xi}}(p',\omega')
  &&=G_0(p,\omega) G_0(p',\omega')
    {\bbox{\mu}}(p,\omega){\bbox{\mu}}(p',\omega')
  \nonumber\\
  &&+G_0(p,\omega) G_0(p',\omega')
    [{\bbox{\mu}}(p,\omega)\bbox{F}^{(0)}(p',\omega')+
     {\bbox{\mu}}(p',\omega')\bbox{F}^{(0)}(p,\omega)]
\end{eqnarray}
   where $\bbox{F}^{(0)}(p,\omega)\equiv\bbox{F}
   [{\bbox{\xi}}^{(0)},p,\omega]$,
   and the term containing two $\bbox{F}^{(0)}(p,\omega)$ has been
   dropped because $\bbox{F}^{(0)}(p,\omega)$ is at least the lowest
   nontrivial order.

   Let
\begin{equation}
   B_N(\tau,\tau')\equiv {1\over N\zeta_0}\sum_p {Q_p^2\over\Lambda_p}
\end{equation}
\begin{equation}
   B_N^B(\tau,t,t')\equiv {2\over N}\sum_p {\cos^2{p\pi\tau\over N}
	      \over\zeta_0\Lambda_p}(1-e^{-\Lambda_p\mid t-t'\mid})
\end{equation}
   We find
\begin{eqnarray}
  \Bigl\langle{\bbox{\xi}}^{(0)}(p,\omega)
   &&\exp[i\sqrt{{2\over N}}
     {{\bf k}}{\bbox{\cdot}}\sum_{p'}\int_{\omega'}e^{-i\omega t'}
     {\bbox{\xi}}(p',\omega')Q_{p'}(\tau,\tau')]\Bigr\rangle=
   \nonumber\\
   &&G_0(p,\omega) G_0(p,-\omega){2\over\zeta_0}\sqrt{{2\over N}}
     i{{\bf k}} e^{i\omega t}Q_p(\tau,\tau')e^{-k^2B_N(\tau,\tau')}
\end{eqnarray}
\begin{eqnarray}
  \Bigl\langle{\bbox{\xi}}^{(0)}(p,\omega){\bbox{\xi}}(p',\omega')
   &&\exp[i\sqrt{{2\over N}}{{\bf k}}{\bbox{\cdot}}\sum_{p''}\int_{\omega''}
     e^{-i\omega'' t}{\bbox{\xi}}^{(0)}(p'',\omega'')Q_{p''}(\tau,\tau')]
     \Bigr\rangle=
   \nonumber\\
   &&G_0(p,\omega) G_0(p,-\omega){2\over\zeta_0}\bigl
     [2\pi\delta(\omega+\omega')
     \delta_{p,p'}\bbox{1}-
   \nonumber\\
   &&G_0(p',\omega') G_0(p',-\omega'){2\over\zeta_0^2}{2\over N}{{\bf k}}
     {{\bf k}}
     e^{i(\omega+\omega')t}Q_p(\tau,\tau')Q_{p'}(\tau,\tau')\bigr]
     e^{-k^2B_N(\tau,\tau')}
\end{eqnarray}
\begin{eqnarray}
  \Bigl\langle{\bbox{\xi}}^{(0)}(p,\omega)
   &&\exp[i\sqrt{{2\over N}}{{\bf k}}{\bbox{\cdot}}
     \sum_{p'}\int_{\omega'}{\bbox{\xi}}(p',\omega')
     \cos{p'\pi\tau\over N}(e^{-i\omega't}-e^{-i\omega't'} )
     \Bigr\rangle=
   \nonumber\\
   &&G_0(p,\omega) G_0(p,-\omega){2\over\zeta_0}\sqrt{{2\over N}}i{{\bf k}}
     (e^{i\omega t}-e^{i\omega t'})\cos{p\pi\tau\over N}
     e^{-k^2B_N^B(\tau,t,t')}
\end{eqnarray}

   As noted in Sec.\ \ref{mode} we confine ourselves to the stationary case.
   Therefore,
\begin{equation}
   \bbox{G}(t,t',{{\bf k}})=\bbox{G}(t-t',{{\bf k}})
\end{equation}
\begin{equation}
   C(t,t',{{\bf k}})=C(t-t',{{\bf k}})
\end{equation}
   where, assuming $\bbox{G}(t,t',1,2)=\bbox{G}(t,t',1-2)$ and
   $C(t,t',1,2)=C(t,t',1-2)$, i.e. translational
   invariance,
\begin{equation}
   \bbox{G}(t,t',{{\bf k}})\equiv\int d{{\bf R}}\bbox{G}(t,t',{{\bf R}})
   e^{-i{{\bf k}}{\bbox{\cdot}}{{\bf R}}}
\end{equation}
\begin{equation}
   C(t,t',{{\bf k}})\equiv\int d{{\bf R}} C(t,t',{{\bf R}})
   e^{-i{{\bf k}}{\bbox{\cdot}}{{\bf R}}}
\end{equation}

   Defining
\begin{equation}
   \bbox{\tilde G}(\tau,\omega,{{\bf k}})\equiv
   \int dt e^{i\omega t}\bbox{G}(t,{{\bf k}})e^{-k^2B_N^B(\tau,t)}
\end{equation}
  ($B_N^B(\tau,t,t')=B_N^B(\tau,t-t')$) and
\begin{equation}
   {\tilde C}(\tau,\omega,{{\bf k}})\equiv
   \int dt e^{i\omega t}C(t,{{\bf k}})e^{-k^2B_N^B(\tau,t)}
\end{equation}
   then ($d$ is the dimensionality)
\begin{eqnarray}
  &&{\langle{\bbox{\xi}}(p,\omega){\bbox{\xi}}(p',\omega')\rangle
   \over
   2\pi\delta(\omega+\omega')G_0(p,\omega)G_0(p,-\omega)}
  \nonumber\\
  &&\qquad
    ={2\over\zeta_0}\delta_{p,p'}\bbox{1}+
      2g_0^2{2\over N}\int d\tau d\tau'\int_{{\bf k}}{1-1/d\over\eta_0 k^2}
      \cos{p\pi\tau\over N}\cos{p'\pi\tau'\over N}e^{-k^2B_N(\tau,\tau')}
      \bbox{1}
  \nonumber\\
  &&\qquad
    +{B^2\over 2\zeta_0}{2\over N}\int d\tau\int_{{\bf k}}{{\bf k}}{{\bf k}}
    \cos{p\pi\tau\over N}\cos{p'\pi\tau'\over N}
    \tilde{C}(\tau,\omega,{{\bf k}})
  \nonumber\\
  &&\qquad
    +{2\over N}{2\over\zeta_0}G_0(p,\omega){\cal R}(p,p',\omega)
    +{2\over N}{2\over\zeta_0}G_0(p',-\omega){\cal R}(p',p,-\omega)
\end{eqnarray}
\begin{eqnarray}
  {\cal R}(p,p',\omega)
    &&\equiv
      {B_0\over\zeta_0}\int d\tau d\tau'\int_{{\bf k}}{k^2\over d}
      \cos{p\pi\tau\over N}Q_{p'}(\tau,\tau')
      e^{-k^2B_N(\tau,\tau')}\bbox{1}
    \nonumber\\
    &&-g_0^2({p'\pi\over N})^2
      \int d\tau d\tau'\int_{{\bf k}}{1-1/d\over\eta_0 k^2}
      \cos{p\pi\tau\over N}\cos{p'\pi\tau'\over N}e^{-k^2B_N(\tau,\tau')}
      \bbox{1}
    \nonumber\\
    &&{B^2\over 2\zeta_0^2}\int d\tau\int_{{\bf k}}
      i[\bbox{\tilde G}(\tau,0,{{\bf k}})-
	\bbox{\tilde G}(\tau,\omega,{{\bf k}})]{\bbox{\cdot}}{{\bf k}}
	{{\bf k}}{{\bf k}}
      \cos{p'\pi\tau\over N}\cos{p\pi\tau\over N}
\end{eqnarray}
   Note that this equation is obtained after higher order terms
   are dropped. Therefore, it does not mean that ${\bbox{\xi}}(p,\omega)$
   is a Gaussian process.

   Define

\begin{equation}
  \delta\Lambda_p^{(1)}=-{B_0\over\zeta_0}{2\over N}\int d\tau d\tau'
       \int_{{{\bf k}}} {k^2\over d}\cos{p\pi\tau\over N}Q_p(\tau,\tau')
       e^{-k^2B_N(\tau,\tau')}
\end{equation}
\begin{equation}
  \delta\Lambda_p^{(2)}=-{g_0^2\over\eta_0}{2\over N}({p\pi\over N})^2
       \int d\tau d\tau' \int_{{\bf k}} {1-1/d\over k^2}
       \cos{p\pi\tau\over N}\cos{p\pi\tau'\over N}
       e^{-k^2B_N(\tau,\tau')}
\end{equation}
\begin{equation}
  \delta\Lambda_p^{(3)}(\omega)=-{B^2\over 2\zeta_0^2}{2\over N}\int d\tau
       \int_{{\bf k}}\cos{p\pi\tau\over N}\cos{p\pi\tau'\over N}
	i[\bbox{\tilde G}(\tau,0,{{\bf k}})-
	  \bbox{\tilde G}(\tau,\omega,{{\bf k}})]
       {\bbox{\cdot}}{{\bf k}}{k^2\over d}
\end{equation}
   The scalar contraction of the tensor product gives (We take $p=p'$
   mode only because of orthogonality of modes.),
\begin{eqnarray}
   \langle{\bbox{\xi}}(p,\omega){\bbox{\cdot}}{\bbox{\xi}}(p,\omega')\rangle
  =&&G_0(p,\omega)G_0(p,-\omega)
   \langle{\bbox{\mu}}(p,\omega){\bbox{\cdot}}{\bbox{\mu}}(p,\omega')\rangle_0
   \nonumber\\
   &&\times[
   1-G_0(p,\omega)(\delta\Lambda_p^{(1)}+\delta\Lambda_p^{(2)}+
		   \delta\Lambda_p^{(3)}(\omega))
   \nonumber\\
   &&\qquad\qquad
    -G_0(p,-\omega)(\delta\Lambda_p^{(1)}+\delta\Lambda_p^{(2)}+
		    \delta\Lambda_p^{(3)}(-\omega))]
\end{eqnarray}
   where $\langle{\bbox{\mu}}(p,\omega){\bbox{\cdot}}
   {\bbox{\mu}}(p,\omega')\rangle_0$ is the value obtained from
   Eq.\ (\ref{mumu-full}) by using zeroth order solution
   Eq.\ (\ref{zeroth}).

   Therefore,
\begin{equation}
   [G_0^{-1}(p,\omega)+\delta\Lambda_p(\omega)]
   [G_0^{-1}(p,-\omega)+\delta\Lambda_p(-\omega)]
   \langle{\bbox{\xi}}(p,\omega){\bbox{\cdot}}{\bbox{\xi}}(p,\omega')\rangle
=\langle{\bbox{\mu}}(p,\omega){\bbox{\cdot}}{\bbox{\mu}}(p,\omega')\rangle_0
\end{equation}
\begin{equation}
   \delta\Lambda_p(\omega)\equiv\delta\Lambda_p^{(1)}+\delta\Lambda_p^{(2)}
		  + \delta\Lambda_p^{(3)}(\omega)
\end{equation}
   The relaxation spectrum is therefore renormalized
\begin{equation}
   \Lambda_p \to \Lambda_p+\delta\Lambda_p(\omega)
\end{equation}
   We note that the first two corrections, $\delta\Lambda_p^{(1)}$ and
   $\delta\Lambda_p^{(1)}$ due to excluded volume effect and hydrodynamic
   effect, respectively, are the same as those found in Ref.\ \cite{spec}.


\begin{references}
\bibitem{super}
   See, for example, D. R. Tilley and J. Tilley, {\it Superfluidity and
	       Superconductivity}, 3rd ed. (Adam Hilger, New York, 1990);
	       D. S. Fisher, M. P. A. Fisher and D. A. Huse, Phys. Rev.
	       {\bf B43}, 130 (1991).
\bibitem{helium}
   S. B. Kim, J. Ma, and M. H. W. Chan, Phys. Rev. Lett. {\bf 71}, 2268
	       (1993); J. Ma, S. B. Kim, L. W. Hrubesh, and M. H. W. Chan,
	       J. Low Temp. Phys. {\bf 93}, 945 (1993); N. Mulders, J. Ma,
	       S. B. Kim, J. S. Yoon, and M. H. W. Chan, J. Low Temp. Phys.
	       {\bf 101}, 95 (1995).
\bibitem{spin}
   See, for example, K. H. Fischer and J. A. Hertz, {\it Spin Glasses}
	       (Cambridge University Press, New York, 1991);
   K. Binder and A. P. Young, Rev. Mod. Phys. {\bf 58}, 801 (1986).
\bibitem{bio}
   D. L. Stein, Proc. Natl. Acad. Sci. USA {\bf 82}, 3670 (1985);
   T. Garel and H. Orland, Europhys. Lett. {\bf 6}, 307 (1988);
   E. I. Shakhnovich and A. M. Gutin, J. Phys. A {\bf 22}, 1647 (1989);
   E. I. Shakhnovich and A. M. Gutin, Biophys. Chem. {\bf 34}, 187 (1989).
\bibitem{replica}
   S. Edwards and P. W. Anderson, J. Phys. F {\bf 5}, 965 (1975).
\bibitem{dyna}
   C. De Dominicis, Phys. Rev. B {\bf 18}, 4913 (1978).
\bibitem{MSR}
   P. C. Martin, E. D. Siggia and H. A. Rose, Phys. Rev. A
		   {\bf 8}, 423 (1973).
   R. V. Jensen, J. Stat. Phys. {\bf 25}, 183 (1981) and
		references therein.
\bibitem{Sompolinsky}
   H. Sompolinsky and A. Zippelius, Phys. Rev. B {\bf 25}, 6860 (1982).
\bibitem{Kirk:p-spin}
   T. R. Kirkpatrick and D. Thirumalai, Phys. Rev. B {\bf 36}, 5388 (1987).
\bibitem{Doi}
   M. Doi and S. F. Edwards, {\it The Theory of Polymer Dynamics}
       (Oxford University Press, Oxford, 1986).
\bibitem{Bird}
   R. B. Bird, C. F. Curtiss, R. C. Armstrong and O. Hassager,
     {\it Dynamics of Polymeric Liquids}, Vol. 2 (Wiley-Interscience,
     New York, 1987).
\bibitem{Oono:review}
   Y. Oono, Adv. Chem. Phys. {\bf 61}, 301 (1985).
\bibitem{Jaga}
   A. Jagannathan, B. Schaub and Y. Oono, Phys. Lett. {\bf A113}, 341 (1985).
\bibitem{Schaub}
   B. Schaub, B. A. Friedman and Y. Oono, Phys. Lett. {\bf A110}, 136 (1985).
\bibitem{vis}
   A. Jagannathan, Y. Oono, and B. Schaub, J. Chem. Phys. {\bf
	     86}, 2276 (1987);
   H. Johannesson and B. Schaub, Phys. Rev. {\bf A35}, 3571 (1987).
\bibitem{Lee-Baldwin-Oono}
   A. Lee, P. R. Baldwin and Y. Oono, Phys. Rev. A {\bf 30}, 968 (1984).
\bibitem{spec}
   D. Jasnow and M. A. Moore, J. Physique Lett. {\bf 38}, L-467 (1977);
   G. F. Al-Noaimi, G. C. Martinez-Mekler and C. A. Wilson,
	      J. Physique Lett. {\bf 39}, L-373 (1978);
   Y. Shiwa and K. Kawasaki, J. Phys. C {\bf 15}, 5345 (1982).
\bibitem{bwkin}
   J. D. Bryngelson and P. Wolynes, J. Phys. Chem. {\bf 93}, 6902 (1989).
\bibitem{Freed:book}
   K. F. Freed, {\it Renormalization Group Theory of Macromolecules}
     (John Wiley and Sons, New York, 1987).
\bibitem{Cloizeaux}
   J. des Cloizeaux and G. Jannink, {\it Polymers in Solution}
      (Oxford University Press, Oxford, 1989).
\bibitem{Oono-Freed:eq}
   Y. Oono and K. F. Freed, J. Chem. Phys. {\bf 75}, 1009 (1981).
\bibitem{Landau}
   L. D. Landau and E. M. Lifshitz, {\it Fluid Mechanics}
	     (Pergamon, New York, 1959).
\bibitem{Oono-Baldwin}
   Y. Oono and P. Baldwin, Phys. Rev. A {\bf 33}, 3391 (1986).
\bibitem{Beenakker}
   C. W. J. Beenakker and P. Mazur, Phys. Fluids, {\bf 28}, 767 (1985)
   discusses a paradox (Smoluchowski paradox)
   caused by this non-applicability.
\bibitem{Oono:aip}
   Y. Oono, in {\it Polymer-Flow Interaction}, AIP Conference Proceeding
     No. 137, edited by Y. Rabin, (American Institute of Physics, New York,
     1985).
\bibitem{Kawasaki-Gunton}
   K. Kawasaki and J. D. Gunton, in {\it Progress in Liquid Physics},
     edited by C. A. Croxton (John Wiley and Sons, 1978).
\bibitem{Vilgis}
   T. A. Vilgis, J. Phys. (France) I {\bf 1}, 1389 (1991).
\bibitem{Schaub:1}
   B. Schaub, D. B. Creamer and H. Johannesson, J. Phys. A {\bf 21}, 1431
			     (1988).
\bibitem{Schaub:2}
   H. Johannesson, D. B. Creamer and B. Schaub, J. Phys. A {\bf 20}, 5071
			     (1987).
\bibitem{Frey}
   E. Frey and D. R. Nelson, J. Phys. (Paris) I, {\bf 1}, 1715 (1991).
\bibitem{Hanggi}
   P. H\"anggi and P. Jung, Adv. Chem. Phys. {\bf 89}, 239 (1995).
\bibitem{Zinn-Justin}
   J. Zinn-Justin, {\it Quantum Field Theory and Critical Phenomena},
      (Oxford University Press, Oxford, 1989).
\bibitem{Bausch}
   R. Bausch, H. K. Jassen and H. Wagner, Z. Phys. B, {\bf 24}, 113 (1976).
\bibitem{Kurchan}
   J. Kurchan, J. Phys. (Paris) I, {\bf 2}, 1333 (1992) attempts to
      formulate spin glass dynamics in terms of supersymmetry.
\bibitem{Das-Mazenko}
   S. P. Das and G. F. Mazenko, Phys. Rev. A, {\bf 34}, 2265 (1986).
\bibitem{McComb}
   W. D. McComb, {\it The Physics of Fluid Turbulence}, (Oxford University
	       Press, Oxford, 1990).
\bibitem{Oono-email}
   Y. Oono (private communication)
\bibitem{Fred}
   However, G. H. Fredrickson and E. Helfand, J. Chem. Phys. {\bf 93},
       2048 (1990) find that this method is useful for the study
       of collective dynamics of polymer solution.
\bibitem{MSR-Jacobian}
   F. Langouche, D. Roekaerts and E. Tirapegui, Physica, A
		{\bf 95}, 252 (1979).
\bibitem{Graham}
   R. Graham, in {\it Springer Tracts in Modern Physics},
		Vol. 66, Springer-Verlag, (1973).
\bibitem{Peliti}
   C. De Dominicis and L. Peliti, Phys. Rev. B {\bf 18}, 353 (1978).
\bibitem{West}
   B. J. West, A. R. Bulsara, K. Lindenberg, V. Seshadri, and K. E. Shuler,
		 Physica, {\bf 97A}, 211 (1979).
\bibitem{Sompolinsky:order}
   H. Sompolinsky, Phys. Rev. Lett. {\bf 47}, 935 (1981).
\bibitem{Zwanzig:gle}
   R. Zwanzig, J. Stat. Phys. {\bf 9}, 215 (1973).
\bibitem{Kawasaki:gle}
   K. Kawasaki, J. Phys. {\bf 6}, 1289 (1973).
\bibitem{Boon}
   J. P. Boon and S. Yip, {\it Molecular Hydrodynamics}
     (Dover, New York, 1991);
   J. P. Hansen and I. R. McDonald, {\it Theory of Simple Liquids}
    2nd ed. (Academic Press, London, 1986);
   D. Forster, {\it Hydrodynamic Fluctuations, Broken Symmetry, and
	   Correlation Functions}, (Addison-Wesley, Reading,
	   Massachusetts, 1990).
\bibitem{Wang}
   M. C. Wang and G. E. Uhlenbeck, Rev. Mod. Phys.
	       Phys. {\bf 17}, 323 (1945).
\bibitem{Zwanzig}
   R. Zwanzig, Phys. Rev. {\bf 124}, 983 (1961).
\bibitem{out-of}
   L. F. Cugliandolo and J. Kurchan, J. Phys. A {\bf 27}, 5749 (1994).
   L. F. Cugliandolo and J. Kurchan, Phil. Mag. B {\bf 71}, 501 (1995).
   L. F. Cugliandolo, J. Kurchan and F. Ritort, Phys. Rev. B {\bf 49},
				   6331 (1994).
   A, Crisanti, H. Horner and H.-J. Sommers, Z. Phys. B {\bf 92}, 257 (1993).
   S. Franz and M. M\'ezard, Europhys. Lett. {\bf 26}, 209 (1994).
   S. Franz and M. M\'ezard, Physica A {\bf 210}, 48 (1994).
   A. Baldassarri, L. F. Cugliandolo, J. Kurchan and G. Parisi,
				     J. Phys. A {\bf 28}, 1831 (1995).
\bibitem{Mori}
   H. Mori and H. Fujisaka, Prog. Theore. Phys. {\bf 49}, 764 (1973).
   K. S. J. Nordholm and R. Zwanzig, J. Stat. Phys. {\bf 11},
	       143 (1974).
\bibitem{Ma}
   S.-K. Ma and G. F. Mazenko, Phys. Rev. {\bf B 11}, 4077 (1975).
\bibitem{Stepanow}
   S. Stepanow, J. Phys. A {\bf 17}, 3041 (1984).
\bibitem{Gutin}
   A. M. Gutin, A. Sali, V. I. Abkevich, M. Karplus, and E. I. Shakhnovich,
       to be published.
\bibitem{deGennes}
   P. G. de Gennes, Phys. Lett. A {\bf 38}, 339 (1972).
\bibitem{Emery-Fisher}
   V. J. Emery, Phys. Rev. B {\bf 11}, 239 (1975);
   D. Jasnow and M. E. Fisher, Phys. Rev., B {\bf 13}, 1112 (1976).
\bibitem{Maritan}
   A. Maritan, M. Cieplak, M. R. Swift, F. Toigo, and J. R. Banavar,
       Phys. Rev. Lett. {\bf 69}, 221 (1992). See also, A. Falicov and
       A. N. Berker, Phys. Rev. Lett. {\bf 74}, 426 (1995).
\bibitem{BEG}
   M. Blume, V. J. Emery, and R. B. Griffiths, Phys. Rev. A {\bf 4}, 1071
       (1971).
\bibitem{Dawson:kin}
   A. Byrne, P. Kiernan, D. Green and K. A. Dawson, J. Chem. Phys. {\bf 102},
		      573 (1995).

\bibitem{ganazzoli}
   F. Ganazzoli, R. La Ferla, and G. Allegra, Macromolecules, {\bf 28},
		      5285 (1995).
\end{references}

\end{document}